\PassOptionsToPackage{numbers,sort&compress}{natbib}
\documentclass[11pt]{article}

\usepackage{geometry, fullpage, authblk}
\usepackage{xr-hyper}
\usepackage{times,enumitem}
\usepackage{hyperref}[]
\hypersetup{
	colorlinks=true,
	linkcolor=blue,
	filecolor=magenta,      
	urlcolor=cyan,
	citecolor=blue,
}
\usepackage{ulem}
\usepackage{url}
\usepackage[f]{esvect}
\usepackage{multirow}
\usepackage{setspace}
\usepackage{lineno}

\usepackage{algorithm}
\usepackage{algorithmic}

\usepackage{bigints}
\usepackage{amsfonts,amsmath,amssymb,amsthm, bbm, bm, mathtools}
\usepackage{wrapfig}
\usepackage{verbatim,float,url}
\usepackage{graphicx}
\usepackage{subcaption}
\usepackage{caption}
\usepackage[round]{natbib}   
\usepackage[english]{babel}
\usepackage{cancel}
\usepackage{color}
\usepackage[dvipsnames]{xcolor}
\usepackage{cleveref}
\usepackage{multirow}
\usepackage{booktabs}
\usepackage{authblk}
\title{This is some thing}
\author[1]{Ziyue Zheng} 
\author[2,3]{Loay J. Jabre} 
\author[2]{Matthew McIlvin} 
\author[2]{Mak A. Saito} 

\author[1]{Sangwon Hyun} 
\affil[1]{University of California, Santa Cruz}

\affil[2]{Marine Chemistry and Geochemistry Department, Woods Hole Oceanographic Institution}
\usepackage[flushleft]{threeparttable}
\usepackage{graphics}
\affil[3]{Department of Biology, Mount Allison University}
\usepackage[flushleft]{threeparttable}
\usepackage{graphics}

\usepackage{tikz}
\usetikzlibrary{automata, arrows.meta, positioning, fit, shapes}

\usepackage{marginnote}

\usepackage{makecell}

\newcommand{\ynew}{y_{\text{new}}}

\newcommand{\ourtitle}{Subcellular proteome niche discovery using semi-supervised functional clustering}

\title{\ourtitle }

\begin{document}

\maketitle

\let\clearpage\relax
\begin{abstract}
Intracellular compartmentalization of proteins underpins their function and the metabolic processes they sustain. Various mass spectrometry–based proteomics methods (subcellular spatial proteomics) now allow high throughput subcellular protein localization. Yet, the curation, analysis and interpretation of these data remain challenging, particularly in non-model organisms where establishing reliable marker proteins is difficult, and in contexts where experimental replication and subcellular fractionation are constrained. Here, we develop \texttt{FSPmix}, a semi-supervised functional clustering method implemented as an open-source R package, which leverages partial annotations from a subset of marker proteins to predict protein subcellular localization from subcellular spatial proteomics data. This method explicitly assumes that protein signatures vary smoothly across subcellular fractions, enabling more robust inference under low signal-to-noise data regimes. We applied \texttt{FSPmix} to a subcellular proteomics dataset from a marine diatom, allowing us to assign probabilistic localizations to proteins and uncover potentially new protein functions. Altogether, this work lays the foundation for more robust statistical analysis and interpretation of subcellular proteomics datasets, particularly in understudied organisms.  
\end{abstract}

\section{Introduction}
\label{sec:intro}

A complex network of biochemical reactions, collectively known as metabolism, sustains all life. These reactions are driven by molecular machines—proteins—that are precisely organized within subcellular niches (e.g. organelles, organelle-like structures, condensates, membranes, etc.) according to their function. Within eukaryotes for example, proteins involved in respiration reside in mitochondria, and photosynthetic proteins localize to chloroplasts. Understanding the subcellular organization of proteins, and their interactions with other proteins, substrates and nano-environments within the cell, is therefore essential to understanding fundamental biological processes. 

\begin{figure}[H]
  \centering
  \includegraphics[width=\linewidth]{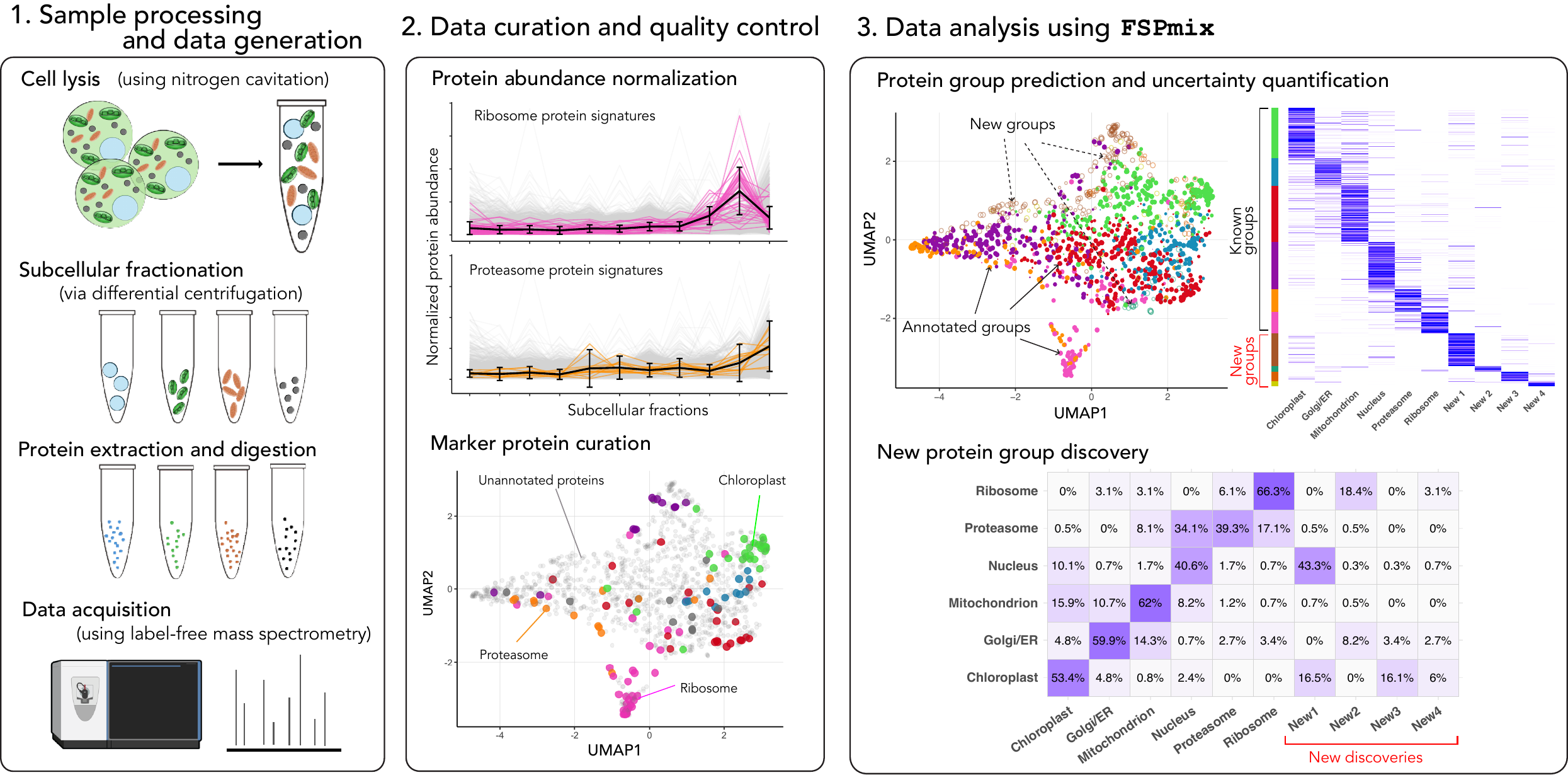}
    \caption{
    \textbf{Workflow of spatial proteomics data analysis using \texttt{FSPmix}.}
   \textbf{(Panel 1)}
   Cells from a marine diatom were lysed and fractionated into 11 subcellular components using differential centrifugation. Fractions were then analyzed by label-free shotgun mass spectrometry. \textbf{(Panel 2)} Abundance data were normalized, then marker proteins from previous studies were used to define six subcellular compartments: Chloroplast, Mitochondrion, Golgi/ER, Nucleus, Ribosome, and Proteasome. In all plots, grey lines or points show unannotated proteins, while the colored lines or points show marker proteins with known memberships. \textbf{(Panel 3)} Our proposed \texttt{FSPmix} model was used to (1) predict group memberships of all proteins, and (2) discover new subcellular protein groups. For more details, see also Figure~\ref{fig:umap-summary}, \ref{fig:responsibility}; for uncertainty quantification, see Figure~\ref{fig:hpb}.}
    \label{fig:intro}
\end{figure}

Subcellular spatial proteomics encompasses a suite of methods such as Dynamic Organellar Maps (DOMs) and Localization of Organelle Proteins by Isotope Tagging (LOPIT)  \citep{LOPIT, schessner2023deep}, which enable high-throughput localization of proteins within cells. These methods typically involve careful cell lysis to break the outer cell membranes while preserving the structural integrity of intracellular niches (Fig~\ref{fig:intro}). The subcellular niches are then separated into fractions, usually by ultracentrifugation, and the proteins within each fraction are measured by mass spectrometry. This produces protein abundance profiles that correspond to various subcellular fractions or niches. Data analysis then relies on a manually curated and annotated subset of well-characterized ``marker proteins" with previously known subcellular localizations, which serve as references to localize unannotated proteins having similar abundance profiles.  

Subcellular spatial proteomics methods been widely applied to diverse systems, including human cells, yeast, Arabidopsis, and Apicomplexan parasites, yielding important insights into cellular organization and fundamental biological processes \citep{nightingale2019protocol,nightingale2019mapping,geladaki2019combining,nikolovski2012putative,barylyuk2020comprehensive}. Despite this, these methods remain particularly difficult to apply to understudied organisms where high-quality marker proteins are scarce, and subcellular niches are not well characterized (e.g.  recent discovery of nitroplast organelle, \cite{coale2024nitrogen}). Furthermore, limited mass spectrometry resources and the lack of established cell lysis and fractionation protocols for these organisms could lead to initially noisy data that are difficult to analyze using available statistical approaches. Here, we directly address these data analysis challenges by introducing Functional Spatial Proteomics Mixture model (FSPmix), a novel semi-supervised statistical mixture model designed to analyze subcellular proteomics datasets with limited marker proteins and low signal-to-noise ratios. 

In statistics and machine learning, semi-supervised learning is a well-established area of research. Pure distance-based methods such as label propagation \citep{zhu2002learning} and label spreading \citep{zhou2003learning} were first designed for semi-supervised learning in computer science fields. These models have high computational efficiency but are prone to overfitting and lack of interpretability. \cite{chung2004difficulties} proposed the semi-supervised approach as a solution to the label-switching problem in mixture models, and proved that even with a small subset of data labeled, the likelihood of the model can have unique maxima. More recently, \cite{kunkel2020anchored} discusses Gaussian mixture models (GMMs) with partially labeled data under a Bayesian framework. Another relevant method area is functional clustering. For example, \cite{huang-mixture} developed a kernel smoothing method to estimate the mixture of Gaussian processes from unsupervised data, with an application in supermarket customer data. Their model focuses on clustering Gaussian processes and assume both mean and variance are smooth across time, while our model assumes only a smooth mean. 

In the subcellular spatial proteomics data analysis literature,  \cite{breckels2013effect} develops a protein group discovery algorithm, phenoDisco, which starts by reducing the dimensions of the data to two dimensions using Principal Component Analysis (PCA), then uses primarily GMMs in a step-wise (one new group every iteration) customized algorithm. \cite{crook2018bayesian} develops a Bayesian mixture model, TAGM, to describe the statistical generative process of the proteins, using $t$-augmented GMMs as a basis. They model the data with a mixture of multiple Gaussian distributions and a single $t$-distribution, to assign data membership to outlier proteins. \cite{crook2020semi-disco} extends this model with a novelty detection procedure (Novelty-TAGM), and estimate the number of novel clusters based on the principle of overfitted mixtures in a Bayesian model. Lastly, \cite{crook2022semi} introduced a non‑parametric Bayesian framework that places a Gaussian process prior on the mean trajectory to capture cross‑fraction correlations, while adopting a homoskedastic assumption that fixes the variance to be constant across all fractions. We will call this method GP-TAGM (note, at the time of writing this article, the software is not available). Table~\ref{tab:method-comparison} summarizes how our method compares to other existing methods. 

\begin{table}[htbp]
  \centering
  \caption{ Feature comparison of spatial proteomics analysis methods.}
  \label{tab:method-comparison}
  \begin{tabular}{lcccccc}
    \toprule
    Method &
    \makecell{Generative\\Model} &
    \makecell{Smooth\\Mean} &
    \makecell{New Group\\Discovery} &
    \makecell{Heteroskedastic\\Variance} &
    \makecell{Designed for\\Low S/N Data} &
    \makecell{Data-driven\\Choice of K} \\
    \midrule
    phenodisco & $\times$ & $\times$ & $\checkmark$ & $\times$ & $\times$ & $\times$ \\
    TAGM       & $\checkmark$ & $\times$ & $\times$ & $\times$ & $\times$ & $\times$ \\
    Novelty-TAGM & $\checkmark$ & $\times$ & $\checkmark$ & $\checkmark$ & $\times$ & $\times$ \\
    GP-TAGM    & $\checkmark$ & $\checkmark$ & $\times$ & $\times$ & $\times$ & $\times$ \\
    FSPmix     & $\checkmark$ & $\checkmark$ & $\checkmark$ & $\checkmark$ & $\checkmark$ & $\checkmark$ \\
    \bottomrule
  \end{tabular}
\end{table}

We develop a semi-supervised cluster model (FSPmix) that is specialized for spatial proteomics data from marine microbes. 
We then apply it to analyze a novel subcellular proteomics dataset
 we will refer to as \textit{Thaps2024}, named after the model diatom {\it Thalassiosira pseudonana} \citep{Armbrust2004}. 
Diatoms are an ecologically and biogeochemically important group of microorganisms, responsible for approximately 20\% of global carbon fixation  \citep{benoiston2017evolution}, yet they remain relatively understudied. We find that FSPmix reliably clusters diatom proteins into subcellular niches, enables discovery of novel protein function, and identifies possible uncharacterized subcellular niches. Moreover, it outperforms existing data analysis methods when benchmarked against our and other comparable datasets, providing a framework for analyzing subcellular spatial proteomics data from relatively understudied organisms. 

Protein signatures in this dataset (Thaps2024) are much harder to separate than popular human cell spatial proteomics datasets in the literature. There are also many fewer signatures than is typical, and only eleven subcellular fractions. For example, the first panel of Fig~\ref{fig:umap-summary} shows a 2-dimensional UMAP (Uniform Manifold Approximation and Projection) visualization of the 2,500 diatom proteins (each consisting of relative abundance measurements across $11$ fractions), where one can clearly see that the clustering structure is weaker than conventional proteomics datasets.  We will refer to this data as having \textit{low signal-to-noise ratio (S/N)} -- in terms of group separability, number of subcellular fractions, and the variability of protein signatures within each group. This low S/N ratio is partially due to resource limitations -- the data is collected (1) without chemical tagging of the components prior to the mass spectrometry, and (2) without taking the ratio of the chemically tagged components, and (3) for a smaller number of subcellular fractions (only eleven). Also, importantly, marker proteins are scarce because marine microbes have not been extensively studied using spatial proteomics technology. This lack of established markers, combined with fewer experimental replicates, leads to greater variability and reduced separability among protein groups.
These factors contribute to the overall low S/N characteristic of the Thaps2024 dataset.

Our model uses two key statistical assumptions: (1) the underlying mean protein signature for each group is \textit{smooth} across fractions, and (2) the variance around the mean is that heteroskedastic and independent. 
These are specific features designed for proteomic data that have a low signal-to-noise ratio. Using these assumptions, our FSPmix allows the user to accurately cluster un-annotated protein signatures, discover new protein groups, and estimate statistical uncertainty of the model estimates.

Lastly, our methods have been implemented in the R programming language in the form of a package called \texttt{FSPmix}. Our package is developed using \texttt{litr} \cite{litr}, a literate programming software for generating R packages, which allows future researchers to easily read and extend the methodology. It also integrates the Thaps2024 dataset and several spatial proteomic datasets originally from the \texttt{pRolocdata} \cite{gatto2014mass}. The package \texttt{FSPmix} and all code to reproduce the numerical experiments in this paper are available at a public repository \cite{FSPmixGit}.



\section{Methods}
\label{sec:data}

\subsection{Marine diatom spatial proteomics dataset description} 

We generated a subcellular proteome dataset from the model diatom \textit{Thalassiosira pseudonana} (CCMP1335) using differential centrifugation, henceforth referred to as Thaps2024 . Briefly, cell cultures were grown in F/2 media and harvested during mid-exponential growth using gentle centrifugation at 4 \textdegree  C. Freshly collected cells were lysed using nitrogen cavitation at 1200 PSI, after which subcellular fractions were separated by differential centrifugation at sequentially increasing spin speeds similar to that of \cite{geladaki2019combining}. A total of 11 fractions were collected. Proteins from each fraction were extracted and digested into peptides, and label-free shotgun measurements were conducted on an Orbitrap Astral mass spectrometer for each fraction. Proteome Discoverer was used for databse seaching, protein inference and quantification.

We manually assigned marker proteins corresponding to six subcellular niches: Chloroplast, Mitochondrion, Golgi/ER,  Nucleus, Ribosome and Proteasome. These assignments were based on previous studies and characterizations in \textit{T. pseudonana} or other phytoplankton species.

\subsection{Statistical method overview}

We model the data to be drawn from a generative statistical mixture model of multivariate Gaussian distributions with means that are smooth. To estimate the parameters of the statistical model, we devise and implement an Expectation-Maximization (EM) algorithm, and also devise a \textit{leave-one-fraction-out} cross-validation method and an information criterion (AIC) for choosing the hyperparameters of the statistical model. Details are described below.

We observe the abundance signatures of each protein across the different subcellular fractions as a $d$–dimensional vector $\boldsymbol{y}_i \in \mathbb R^d$. For a given protein, the $j$-th entry $y_{ij}$ (for $j=1,\dots d$) denotes the relative abundance of protein $i$ measured in the $j$-th  fraction. The fraction indices $j=1,\cdots, d$ are ordered so that the larger indices correspond to progressively faster centrifugal speeds. We observe $n$ such proteins $\boldsymbol y_1, \cdots, \boldsymbol y_n$ that are indexed by $i=1,...,n$.

Every protein $\boldsymbol y_{i}$ is thought to belong to a single group, $z_i=k$, where $k$ is a group index that is a positive integer. In our data, $k$ corresponds to a type of protein in a biologically defined subcellular niche; for instance, Chloroplast or Ribosome. However, only a subset of all measured proteins have been assigned to a known group (one of $K$ groups); we refer to such a protein as being \textit{annotated}. 
Furthermore, we assume it is possible that there are \textit{unidentified} protein groups in the data. Specifically, we assume the true number of protein groups in the data is $K$, while only a smaller number of $\tilde{K}$ groups have partial labels, and there are $K_0 = K-\tilde K$ unidentified protein groups in the data.  

Next, we use $\mathcal{L}_k$ to denote the non-overlapping sets of protein indices assigned to group $k$, i.e., $\mathcal{L}_k = \{i:z_i=k\}$, whose union set is $ \mathcal{L} = \cup_{k=1}^{\tilde{K}} \mathcal{L}_k$ and whose complement set is
$\mathcal{U}=\{1,\dots,n\} \setminus \mathcal{L}$. Using this notation, $\{\boldsymbol{y}_i, i \in \mathcal{L}\}$ are annotated proteins with known group membership $z_i$, while the rest $\{\boldsymbol{y}_i, i \in \mathcal{U}\}$ are unlabeled proteins with unknown $z_i$.

In this paper, we devise a methodology to address three data analysis problems:
\begin{enumerate}
    \item \textbf{Estimate group membership for unlabeled data.} Estimate the group membership probabilities for all unlabeled data: $P(z_i = k)$ for $k = 1,\dots,K$ for all unidentified proteins $\{y_i:i \in \mathcal{U}\}$. (Section~\ref{sec:prob-estimation}).
    \item \textbf{Cluster-conditional protein signature prediction.} Characterize regions in the space of protein curves where a newly observed profile $\boldsymbol{y}_{\text{new}}$ is most likely to belong to group $k$, and construct a prediction band across the different subcellular fractions (Section~\ref{sec:high-prob-region}).
    \item \textbf{New protein group discovery.} Discover the number of new unidentified latent protein groups (or subcellular niches) $K_0 = K - \tilde{K} $ and estimate group membership probabilities $P(z_i=k)$ for $k=1,...,K$ (Section~\ref{sec:selectK0}).
\end{enumerate}

In Section~\ref{sec:simulations}, we conduct extensive out-of-sample validation of our method, and compare it with existing methods most similar with ours, and show that our method has more robust validation performance due to specific modeling choices. In Section~\ref{sec:real-data}, we apply our method to a  marine phytoplankton spatial proteomics dataset (Thaps2024), where we make several new group discoveries and discuss some biological implications.


\subsection{Semi-supervised functional clustering model and EM algorithm}

\subsubsection{Statistical model for protein signatures}


We first devise a statistical generative model that describes how the protein signatures may have been formed. Protein signatures $\{\boldsymbol{y}_i\}_{i=1}^n$ are thought of as independent and identically distributed samples from a probabilistic mixture of $K$ different distributions centered around group means $\{\boldsymbol{\mu}_k\}_{k=1}^K$. For each protein $\boldsymbol{y}_i$, the latent variable $z_i$ determines the group membership of that protein, and the probabilistic distribution of $z_i$ is:
\begin{equation*}
P(z_i=k)=\pi_k\; \;\;\text{for}\;\; k=1,\cdots, K,
\end{equation*}
and furthermore, given $z_i=k$, the data $y_i$ is observed as a noisy protein signature around the mean signature $\mu_k$,
\begin{equation*}
\left( \boldsymbol{y}_{i} \mid z_i = k \right) \sim \mathcal{N}_{d}(\boldsymbol{\mu}_k, \boldsymbol{\Sigma}_{k}),
\end{equation*}
where $\boldsymbol{\Sigma}_k$ is the covariance matrix of the $k$ th group across all fractions.

A visual inspection of the original protein's signatures $y_{ij}$ plainly reveals the values to be correlated across the subcellular fractions $j=1,\cdots,d$. Thus, we make the useful assumption that the protein signature's underlying group means $\{\boldsymbol \mu_k\}_{k=1}^K$ are smoothly varying across subcellular fractions. That is, the $k$-th mean at the $j$-th fraction is: 
\begin{equation}
\label{eq:fpc}
    \boldsymbol \mu_{kj} = \sum_{q=1}^{\infty} \xi_{kq} \nu_{kqj},
\end{equation}
where $\xi_{kq}$ are functional principal component scores and $\nu_{kqj}$ are the eigenfunctions evaluated at $j$. We note that \cite{yao2003shrinkage} used kernel-based smoothing methods to estimate the smoothed mean function in a similar setting. Our model for estimating the mean functions also uses kernel smoothing method to capture their smooth trajectories. 

Next, we will assume that the correlation between subcellular fractions $j$ are all captured by the mean $\boldsymbol \mu_k$. Thus, the covariance matrix $\Sigma_k$ of $y_i$ around $\mu_k$ is a diagonal matrix:
\begin{equation}
\label{eq:sigma}
\boldsymbol{\Sigma}_k = 
\begin{pmatrix}
\sigma_{k1}^2 & 0 & \dots & 0 \\
0 & \sigma_{k2}^2 & \dots & 0 \\
\vdots & \vdots & \ddots & \vdots \\
0 & 0 & \dots & \sigma_{kd}^2
\end{pmatrix},
\end{equation}
where the heteroskedastic variances $\sigma_{kj}^2$ of the data at each fraction $j$ are free parameters.
This key statistical assumption allows the most efficient statistical modeling of the problem. A more complicated and flexible model for the $d$-by-$d$ matrix $\Sigma_k$
may contain up to $Kd^2/2$ parameters, which can be infeasible to estimate with limited data. Also, a simpler model that assumes the variance to be homoskedastic ($\sigma_{k1}^2 = \cdots, \sigma_{kd}^2$) is inaccurate, and will only perform clustering reasonably for proteomic data with a strong signal-to-noise ratio \citep{crook2022semi}. Assuming heteroskedastic noise is thus apt for statistically efficient modeling of our low S/N diatom dataset, and is motivated from plain observation of the protein signatures' noise around their group means in the data.

\subsubsection{Expectation-Maximization algorithm}
The model parameters to estimate are $\theta = ([\{\mu_{kj}\}, \{\sigma_{kj}\}, \{\pi_{k}\}]_{j=1}^{d})_{k=1}^K$, which describe the mean, variance and relative probability of the protein signatures in each group. In order to estimate the model parameters using maximum likelihood, we devise an Expectation-Maximization (EM) algorithm.  We describe this next.

The statistical likelihood of the model is the joint mixture likelihood of every protein signature $y_i, i=1,\cdots, n$:
\begin{equation}\label{eq:log-likelihood}
L(\theta|y) = \prod_{i=1}^n  \left[ \sum_{k=1}^K  \pi_{k} \prod_{j=1}^d N(y_{ij}; \mu_{kj}, \sigma_{kj}^2)\right]
\end{equation}
where $N(\cdot;a,b)$ denotes the normal density with mean $a$ and variance $b$, and where the $\pi_{k}$ denotes the group probability $P(z_i = k)$. Next, introducing the {\it latent} variable $\{z_i: i = 1,\cdots, n\}$, this becomes
\begin{equation}\label{eq:likelihood-with-latent}
L(\theta|y) = \prod_{i=1}^n   \prod_{k=1}^K \left[  \pi_{k}  \prod_{j=1}^d N(y_{ij}; \mu_{kj}, \sigma_{kj}^2)\right]^{\mathbbm{1}(z_i = k)}.
\end{equation}
Taking the natural logarithm, this becomes the log-likelihood:
\begin{equation}\label{eq:loglikelihood-with-latent}
log(L(\theta|y)) = \sum_{i=1}^n   \sum_{k=1}^K {\mathbbm{1}(z_i = k)}\log \left[  \pi_{k}  \prod_{j=1}^d N(y_{ij}; \mu_{kj}, \sigma_{kj}^2)\right].
\end{equation}
The conditional expectation of this log likelihood, i.e., with respect to $z_i | y_{i}$, and for a set of parameters $\tilde \theta$ is:
\begin{equation}\label{eq:complete-log-likelihood}
Q_{\tilde \theta}(\theta | y) = \sum_{i=1}^n  \sum_{k=1}^K  \gamma_{ik}
\log  \left[ \pi_{k} \prod_{j=1}^d    N(y_{ij}; \mu_{kj}, \sigma_{kj}^2)\right].
\end{equation}
where we have introduced the conditional probability of membership $\gamma_{ik} = P_{\tilde \theta}(z_i=k|y_i),$ itself a conditional expectation of $\mathbbm{1}(z_i=k)$ assuming $\tilde \theta$. This is sometimes called ``responsibilities'' in the literature. We will aim to iteratively maximize the surrogate likelihood \eqref{eq:complete-log-likelihood} instead of \eqref{eq:log-likelihood}, since the latter is non-convex as a function of $\theta$ and thus more difficult to directly maximize. A majorization-minimization algorithm like an EM algorithm will make nonzero gains in the likelihood over iterations, and is thus an effective strategy for maximum likelihood estimation in similar nonconvex problems \cite{Dempster1977}. Also note if the protein $i$ has already been annotated -- so that $i\in \mathcal L_k$ for a known group $k$ -- then the membership probability $\gamma_{il}$ takes value $1$ if $l=k$, and $0$ otherwise.


The maximization with respect to $\mu$ in \eqref{eq:fpc} can be thought of as a maximization with respect to $\xi$, and can be approximated using an appropriate basis function $\nu$ (e.g., Fourier basis \citep{sollich2004using}) penalizing the likelihood by adding a penalty on the coefficients $\xi$. However, choosing basis functions might be ineffective when the data's true structure is not well-represented by that basis, and requires extra regularization of basis coefficients. Alternatively, we use a kernel-smoothing approach, which estimates directly from data without assuming a global basis, to approximate the mean function, which has the following form at the $k$-th group and the $t$-th fraction:
\begin{equation} \label{eq:mstep-mu}
\hat \mu_{kt} = \frac{\sum_{i=1}^n \sum_{j=1}^d w_{ijkt} y_{ij}}{\sum_{i=1}^n \sum_{j=1}^d w_{ijkt}},
\end{equation}
where $w_{ijkt} = \gamma_{ik} K_h(t , j)$. For the kernel function $K_h(\cdot,\cdot)$, we use the Radial Basis Function (RBF) kernel,
\begin{equation*}
    K_h(t,j) = \exp(-\frac{|t-j|^2}{2h}),
\end{equation*}
noting that the choice of the bandwidth $h$ is much more consequential \citep{wasserman2006all} than the choice of the functional form of the kernel. In Section~\ref{sec:LOFO} we develop and test a cross-validation scheme to choose an optimal bandwidth $h$.  Also, from visual inspection of the data, we see that the variability of protein signatures around their group means to be quite heterogeneous but without correlation across fractions $j=1,\cdots, d$.
Thus, we model the variance component  $\{\sigma_{kj}^2, j = 1,\cdots, d\}$ to {\it not} be smooth, and instead, are assumed to be heteroskedastic -- separate and independent across subcellular fractions $j=1,\cdots,d$ (in contrast to a similar model proposed in \cite{huang-mixture}, which takes both the mean and the variance to be smooth).

The full EM algorithm is to initialize parameters (step 1), then cycle through steps 2-5:
\begin{enumerate}

\item Parameter Initialization: For each group in $k \in \{ 1,...,K\}$ with annotated proteins, draw a subset $\mathcal{S}_k \subset \mathcal{L}_k$ (of size $|\mathcal{L}_k|/2$). Also, for each group $k \in \{K+1, \cdots, K+K_0\}$ with no annotations, draw a random subset $\mathcal{S}_k$ of $\mathcal{U}_k$ (of size $0.01 \cdot|\mathcal{U}_k|$), and initialize the mean and variances as follows:
\begin{align*}
\mu_{k}^{(0)} &= \frac{1}{|\mathcal{S}_k|}\sum_{i \in \mathcal{S}_k} y_{i}\\
\sigma_{kj}^{2^{(0)}}  &= \frac{1}{|\mathcal{S}_k|-1}\sum_{i \in \mathcal{S}_k} (y_{i}-\mu_{k}^{(0)})^2 \;\text{ for all } j=1,\cdots, d.
\end{align*}
The choice of the sizes of $|\mathcal{S}_k|$ ensures that the initial choices of parameters capture diverse potential protein signature structures from both types of groups, labeled and unlabeled. 


\item E step: Update the responsibilities given the latest model estimates 
\begin{equation}
\label{eq:estep}
\hat \gamma_{ik}  = \begin{cases}
\frac{\hat{\pi}_k \prod_{j=1}^d f(y_{ij}; \hat{\mu}_{kj}, \hat{\sigma}^2_{kj})}{\sum_{k=1}^K \hat{\pi}_k \prod_{j=1}^d f(y_{ij}; \hat{\mu}_{kj}, \hat{\sigma}^2_{kj})} & \text{if } i \in \mathcal{U} \\
\; 1 & \text{if } i \in \mathcal{L}_k \\
\; 0 & \text{if } i \in \mathcal{L} \setminus \mathcal{L}_k
\end{cases}
\end{equation}
Also update the cluster probability as
\begin{equation*}
    \hat{\pi}_{k} = \frac{1}{n} \sum_{i=1}^n \hat{\gamma}_{ik}.
\end{equation*}
\item M step for $\mu$: Update the mean estimates by the smooth weighted average, as in \eqref{eq:mstep-mu},
\begin{equation*}
    \hat \mu_{kt} = \frac{\sum_{i=1}^n \sum_{j=1}^d \hat{\gamma}_{ik} K_h(t , j) y_{ij}}{\sum_{i=1}^n \sum_{j=1}^d \hat{\gamma}_{ik} K_h(t , j)}
\end{equation*}
\item M step for $\sigma$: Update the variance estimates by the weighted sample variance
\begin{equation*}
\hat \sigma_{kt}^2 = \frac{\sum_{i=1}^n \hat{\gamma}_{ik} (y_{it} - \hat{\mu}_{kt})^2}{\sum_{i=1}^n \hat{\gamma}_{ik}}.
\end{equation*}
\end{enumerate}

Repeat steps 2-4 until the objective value in \eqref{eq:log-likelihood} converges numerically. Also, repeatedly run the entire EM algorithm multiple times with random initial values from step 1 for a better chance at obtaining the global optimum.

\subsection{Predicting membership probabilities of new proteins}
\label{sec:prob-estimation}

Given a training dataset and a new protein $y_{\text{new}}$, we can predict the $K$ membership probabilities from our model estimated on the training dataset as the conditional probabilities, for $k = 1,2,\cdots, K$:
\begin{equation}
\label{eq:new-responsibility}
    \hat \gamma_k (\ynew) = P\left(z_{\text{new}} = k \mid \{\hat\mu_k\}_k, \{\hat \sigma_{k,j}\}_{k,j}, \{\hat\pi_{k}\}_{k} \right),
\end{equation}
which use trained FSPmix parameters $\{\hat\mu_k\}_k, \{\hat \sigma_{k,j}\}_{k,j}$ and $\{\hat\pi_{k}\}_{k}$. The probabilities in \eqref{eq:new-responsibility} are calculated directly as:
\begin{equation}
\label{eq:new-responsibility}
\hat \gamma_k(\ynew) = \frac{\hat \pi_k \prod_{j=1}^d f(\ynew; \hat \mu_{kj}, \hat \sigma^2_{kj})}{\sum_{k=1}^{K} \hat \pi_k \prod_{j=1}^d f(\ynew; \hat \mu_{kj}, \hat \sigma^2_{kj})}.
\end{equation}
Next, estimating a single membership of a new protein -- one integer between $1$ and $K$ -- can be done in two ways. First, a {\it hard} estimate of a cluster membership can be made simply by taking the maximizer of $\{\hat \gamma_k(y_{\text{new}})\}_{k=1}^K$. This has the simple interpretation as the most likely cluster membership given the training data and results in prediction regions that will unnaturally {\it partition} the $d$-dimensional protein signature space. On the other hand, a {\it soft} estimate of a cluster membership is made by a single random multinomial  draw between $1$ and $K$ according to the cluster probabilities:
\begin{equation}
\label{eq:coinflip}
\hat k^{\text{soft}}(\ynew) \sim \text{Multinom}\left(n=1,\; \{\hat \gamma_k(\ynew)\}_k\right).
\end{equation}
In our paper, we opt for the soft, estimate \eqref{eq:coinflip} of new protein memberships based on the parameter values from the estimated FSPmix model.


\subsection{Prediction bands for protein groups}
\label{sec:high-prob-region}

For each of the $K$ protein groups, we can also form a {\it prediction band}, in protein signature space, which is highly likely to contain new protein signatures of that group. 
Such a prediction band plays a dual role in our pipeline:  
(1) they visualize the characteristic variability of each protein group, helping scientists judge whether a measured protein has a profile consistent with group $k$; and  
(2) they formalize uncertainty, offering statistically principled tolerance regions that remain valid even when the underlying cluster
assignments of the dataset are only partially known (that is, in the semi–supervised setting).
These properties make prediction bands a valuable tool for both exploratory analysis and downstream decision-making.

\subsubsection{Target coverage of prediction bands}
\label{sec:target-cover}

Let $y^{\text{new}} \in \mathbb{R}^d$ denote the observed profile for a new protein and let $y^{\text{new}}_j$ be its abundance at fraction $j$. Denote $z\in\{1,\dots,K\} \text{, where } K=\tilde{K}+K_0$, as the latent group membership of this protein, corresponding to either one of the $\tilde K$ labeled groups or the $K_0$ new groups. Our goal is to construct, for every cluster $k$, a data-driven set
\begin{equation*}
C_k \;=\;
\Bigl\{\,a \in \mathbb{R}^d :
C_{kj}^{\mathrm{lower}}\le a_j \le C_{kj}^{\mathrm{upper}}
\text{ for all } j=1,\dots,m \Bigr\},
\end{equation*}
that has \textit{simultaneous} coverage over all fractions $j$:
\begin{equation}
  \Pr \bigl\{ y^{\text{new}} \in C_k \mid z=k\bigr \}
  \;\ge\;1-\alpha,
  \label{eq:globalCoverage}
\end{equation}
i.e., $C_k$ is an $(1-\alpha)$ probability simultaneous band for the signature of a new protein belonging to group $k$.

However, the $K_0$ new groups are not accompanied by any labeled proteins in the training dataset, so the likelihood is invariant to any permutation of their labels. Hence, when one re‑estimates a model by maximizing the likelihood -- with different data, or even with the same data -- the new group indexed by $k>\tilde{K}$ in one estimated model need not correspond to the group carrying the same index in another estimated model. This is known as the label-switching problem. In our experiments, we relabel the parameters of the $K_0$ new groups $\tilde{\theta} = \{ \mu_k, \sigma_k, \pi_k \}_{k=\tilde{K} + 1, \dots, K}$ by reordering them according to group mean $\frac{1}{d}\sum_{j=1}^d \mu_{kj}$, from low to high. 
The coverage statement in \eqref{eq:globalCoverage} for new groups becomes
\begin{equation} 
\label{eq:globalCoverage-after-ordering}
\Pr \bigl\{ y^{\text{new}} \in C_k \mid z=k(\tilde{\theta}) , \tilde{\theta} \bigr \} \ge\;1-\alpha,
\end{equation}
where $k(\tilde{\theta})$ denotes the rank-ordered index after this relabeling. 
This statement means that $C_k$ simultaneously covers the entire protein signature from the $k$-th group with probability $1-\alpha$, given the sorted groups of protein groups. Equation \eqref{eq:globalCoverage-after-ordering} will be the target coverage of $C_k$.

\subsubsection{Prediction band construction}

In order to estimate more realistic distributions of protein signatures in forming prediction bands $C_k$, we use non-parametrically estimated distributions at each fraction $j$ instead of the Gaussian distributions that we assumed in our original model.  First, take a subcollection of protein curves $\{y^*_{ik\cdot}, i = 1,\cdots, \hat n_k\}$ from the training data that are deemed to be in group $k$ according to a multinomial membership draw (as in \eqref{eq:coinflip}) from the estimated FSPmix model. Then, we estimate the probability density $\hat f_{kj}(\cdot)$ using a kernel density estimator \cite{bw1986density} at each fraction $j$ as:
\begin{equation*}
    \hat{f}_{kj}(x) = \frac{1}{n_k}\sum_{i=1}^{n_k} \frac{1}{h_{kj}} \cdot K_{h_{kj}}(x,y^*_{ikj})
\end{equation*}
which uses a standard radial basis function (RBF) kernel function $K_h(a,b) = \exp\left(-\frac{|a-b|^2}{2h}\right)$. 
For choosing the bandwidth $h_{kj}$, we use the normal reference rule introduced by \cite{scott1992multivariate}. Specifically, for the $j$-th fraction in the $k$-th group, 
we use $h_{kj} = 1.06\,\hat{\sigma}_{kj}\, \hat{n}_k^{-1/5}$
where $\hat{\sigma}_{kj}$ is the estimated standard deviation of the $k$-th group at the $j$-th fraction, and $\hat{n}_k$ is the number of proteins predicted to belong to group $k$.

Next, for each fraction $j$ in group $k$, let us call $\hat F_{kj}(\cdot)$ the estimated cumulative distribution function (CDF) of $y$ which is the cumulative integral of $\hat f_{kj}(\cdot)$:
\begin{equation*}
    \hat F_{kj}(x) = \int_{-\infty}^x \hat f_{kj}(u)\;du.
\end{equation*}
We then define our prediction band for group $k$ at each channel $j$ to be between:
\begin{equation}
  C_{kj}^{\mathrm{lower}}
  \;=\;
  \hat F_{kj}^{-1}\!\bigl(\alpha^\star / 2\bigr),
  \qquad
  C_{kj}^{\mathrm{upper}}
  \;=\;
  \hat F_{kj}^{-1}\!\bigl(1-\alpha^\star/2\bigr),
  \label{eq:marginalLimits}
\end{equation}
lower and upper bounds of a $(1-\alpha^\star)$ probability interval of $\hat F_{kj}(\cdot)$. We furthermore choose $\alpha^\star = \alpha / d$, a Bonferroni correction \cite{miller1981simultaneous} of $\alpha$ to aim at \textit{simultaneous} coverage as stated in \eqref{eq:globalCoverage-after-ordering}, so that the probability of a new protein's signature being fully contained within its corresponding prediction band is at least $1-\alpha$.

\begin{figure}[H]
    \centering
    \includegraphics[width=0.8\linewidth]{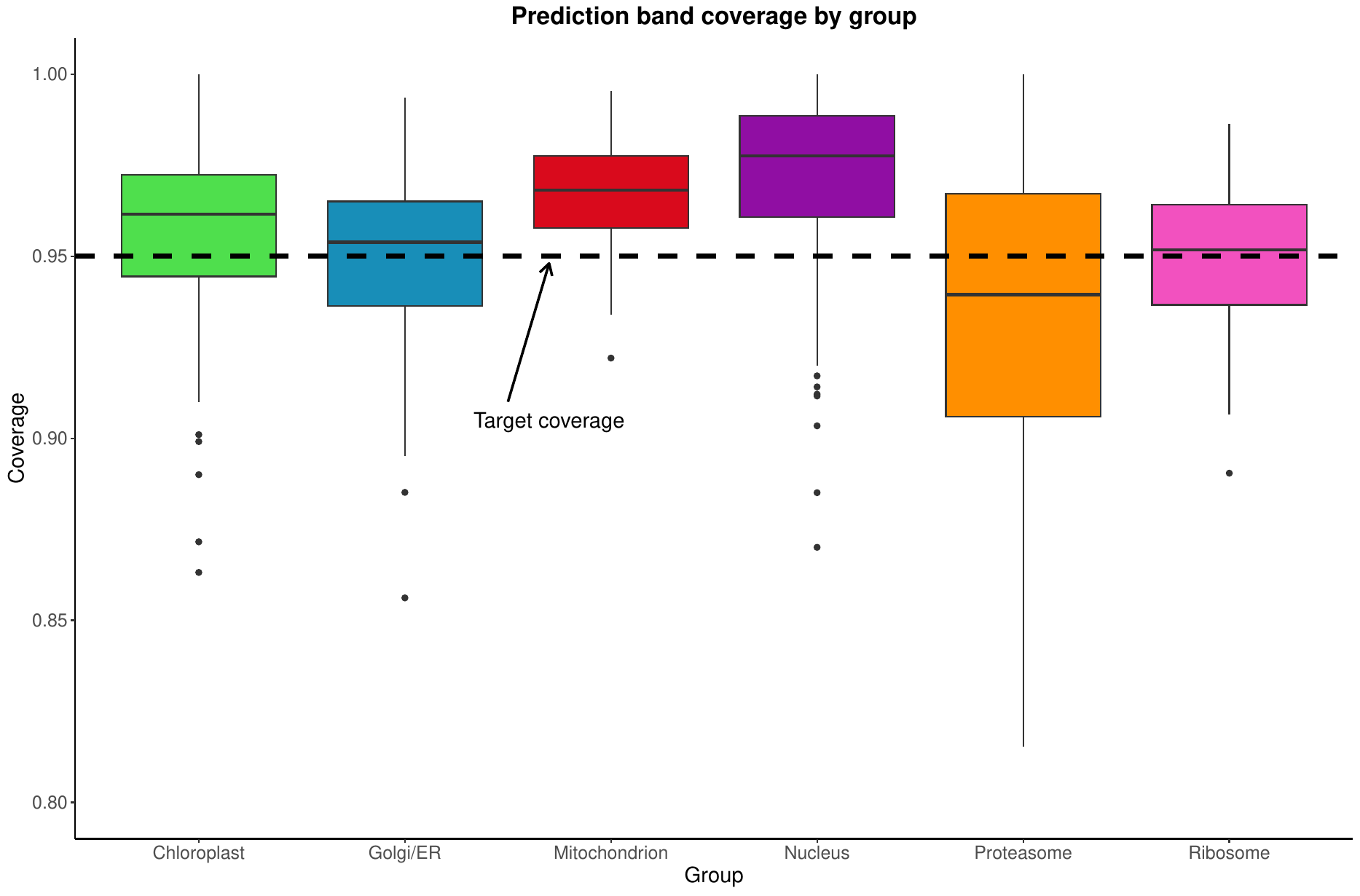}
    \caption{
    \textbf{Coverage of prediction bands in a resampling-based simulation experiment.} The boxplots show the empirical coverage of prediction bands for each protein group, calculated from a pseudo-real simulation (described in Section~\ref{sec:high-prob-region}). In this experiment, we first generated predicted labels using an FSPmix model applied to the full dataset, split it into a training and validation set, then calculated the proportion of proteins in the validation set whose profiles are entirely contained within the correct prediction bands produced from the training set. The horizontal dashed line indicates the target coverage rate of 95\%; all six groups achieve (or nearly achieve) the 95\% simultaneous coverage, on average.}
    \label{fig:hpb_cover}
\end{figure}

To demonstrate that our prediction bands achieve near target coverage, we designed a simulated experiment. First, we fit an FSPmix model to the entire Thaps2024 dataset to obtain predicted labels for every protein using the method in Section \ref{sec:prob-estimation}. Then we fixed these labels as ground truth and performed a resampling procedure in which we repeatedly sampled the data into two independent subsets, each comprising 60\% of the full dataset: one for training and one for validation. The training dataset (with only 20\% of the labels intact, in order to mimic the relative amount of partial labels in the real dataset) was used to construct prediction bands, by fitting another FSPmix model on this dataset and then using \eqref{eq:marginalLimits} with target coverage of $\alpha=0.95$. Next, all proteins in the validation dataset were used to calculate how often their signatures were fully contained within the prediction bands corresponding to their labels. Fig \ref{fig:hpb_cover} shows the results of this experiment. The boxplots indicate that all six groups achieve (or nearly achieve) the target 95\% simultaneous coverage.

\subsection{Model selection}
Our proposed FSPmix model is governed by a small set of hyperparameters that control the smoothness of the proteins' mean trajectories and the number of the latent groups.
Because these choices can impact predictive accuracy and biological interpretability, we adopt a data-driven tuning strategy rather than fixing them a priori.
Section~\ref{sec:LOFO} introduces a cross-validation criterion for selecting
the kernel bandwidth $h$ that regulates smoothness, and
Section~\ref{sec:selectK0} describes an information-theoretic approach for determining the number of additional (previously unlabeled) clusters $K_0$. 
Furthermore, we adopt a greedy search over $h$ and $K_0$ -- we first fix a relatively large $K_0$ (about 2$\tilde{K}$) to select the optimal $h$, and subsequently select $K_0$ from nonnegative integer values.

\subsubsection{Choosing bandwidth $h$}
\label{sec:LOFO}





In nonparametric statistics, leave-one-out cross-validation (LOO-CV) \cite{wasserman2006all} is a commonly used technique to choose the bandwidth $h$ that minimizes the prediction error by holding out one observation at a time. However, applying LOO-CV in our setting would require leaving out every protein at every fraction, requiring the model to be estimated $n\times d$ times, which is computationally infeasible in practice. As an approximate strategy, we adapt this idea to functional mixture models and develop a leave-one-fraction-out cross-validation (LOFO-CV) procedure, which leaves out an entire fraction at a time and thus only requires fitting the model $d$ times.  We describe the procedure next.



For a certain fraction $j \in \{1,2,\cdots, d\}$, denote by $Y^{(-j)}$ the held-in data $\{ y_{ij'} \}_{i=1, \cdots, n, j' \neq j}$, and denote by $Y^{(j)}$ the held-out data. Also, denote the membership probability estimate of protein curve $i$ as $\hat{\gamma}_{ik}^{(-j)}$, computed using the model estimated from the held-in data $Y^{(-j)}$.
We will use $\hat \gamma_{ik}^{(-j)}$ as a proxy for the true membership probabilities $\gamma_{ik} = \mathbb{P}(z_i = k)$ of the $i$'th curve, implicitly assuming that the true membership $\gamma_{ik}$ can be reasonably estimated using only $Y^{(-j)}$. Next, notice that the true group mean at fraction $j$ can be approximated as:
\begin{equation}
\label{eq:extrapolated-mean}
\tilde{\mu}_{kj} = \sum_{i=1}^{n} \hat{\gamma}_{ik}^{(-j)} y_{ij},
\end{equation}
a weighted average of $Y^{(j)}$ which we will treat as the estimation target. A smoothed prediction of \eqref{eq:extrapolated-mean} can be made using the model trained on $Y^{(-j)}$ as a kernel-weighted average :
\begin{equation*}
\hat{\mu}_{kj}^{(-j)} = \frac{ \sum_{j'=1, j' \neq j}^d K_h(j',j) \hat{\mu}_{kj'}^{(-j)} }{ \sum_{j'=1, j' \neq j}^d K_h(j',j) }.
\end{equation*}
Similarly define the extrapolated data noise level $\tilde{\sigma}_{kj}^2$ of group $k$ at fraction $j$ as:
\begin{equation*}
\tilde \sigma_{kj}^2 = \sum_{i=1}^{n} \hat{\gamma}_{ik}^{(-j)} (y_{ij} - \tilde \mu_{kj})^2.
\end{equation*}
We measure the error incurred by using $\hat{\mu}_{kj}^{(-j)}$ to estimate $\tilde{\mu}_{kj}$ as $\frac{1}{ \tilde{\sigma}_{kj}^2/n_k } \cdot\left( \tilde{\mu}_{kj} - \hat{\mu}_{kj}^{(-j)} \right)^2 $, which includes a normalization by an appropriately scaled variance estimate $\tilde{\sigma}_{kj}^2/n_k$ for $n_k=\sum_{i=1}^n \hat{\gamma}_{ki}^{(-j)}$. Next, averaging the error over all groups $k$ and channels $j$, we arrive at the following leave-one-fraction-out (LOFO) cross-validation score:
\begin{equation}
    \label{eq:cvloss}
    \text{CV}(h) = \frac{1}{d} \sum_{j=1}^{d} \sum_{k=1}^{K}\hat \pi_k^{(-j)} \frac{ \left( \tilde{\mu}_{kj} - \hat{\mu}_{kj}^{(-j)} \right)^2 }{ \tilde{\sigma}_{kj}^2/n_k },
\end{equation}
We propose choosing $h$ as the value that minimizes the model's average prediction error $\text{CV}(h)$ in \eqref{eq:cvloss}, using a grid search over $h$.

The right-hand-side panels of Fig \ref{fig:validating_choosing_hyper} compare  the performance of two criteria for selecting the bandwidth $h$. 
Panel B1 shows the corresponding LOFO-CV score -- the main tuning strategy we propose -- while panel B2 shows the hold-out prediction accuracy of group memberships on a grid of candidate values $h$. Comparing the two panels, we observe that the LOFO-CV scores are less variable than the membership prediction accuracies. This is due to the sensitivity of the latter approach to the specific protein curves selected or excluded during the repeated train-test procedure -- requiring more replicates. The former LOFO-CV approach eschews this issue. But notably, the optimal $h$ chosen from LOFO-CV scores in panel B1 aligns closely with the optimal $h$ chosen from prediction accuracy in panel B2, indicating strong agreement between the two bandwidth selection approaches.


\subsubsection{Choosing number of new clusters $K_0$}
\label{sec:selectK0}
Recall that for the data points $i \in\mathcal{L}$, we have pre-existing labels $k \in\{ 1,\cdots, \tilde K\}$.  However, the true number of groups in our dataset is likely to be larger than $\tilde K$. We assume there are $K_0$ clusters without priori membership labels so that the total number of groups is $\tilde K+K_0$. This section describes how we use a data-driven criterion to choose the number of undiscovered protein groups $K_0$.

Determining how many clusters are present in unlabeled data is a long–standing problem with no single solution. Distance–based validity criteria such as the Dunn index, silhouette width, or the ratio of between- to within-cluster dispersion reward well-separated partitions; information criteria (AIC, BIC, ICL) balance goodness-of-fit against model complexity, while resampling methods (e.g. stability analysis) emphasize reproducibility.  Because our primary goal is accurate estimation of a statistical generative model for protein profiles, we adopt the Akaike Information Criterion (AIC):
\begin{equation}
    \text{AIC}(K_0) = 2(K_0 + \tilde{K})d - 2\ln(\hat{L}(K_0)),
    \label{eq:AIC}
\end{equation}
where $\hat{L}(K_0)$ is the data likelihood of the current model, $K_0+\tilde{K}$ is the number of total clusters and $d$ is number of subcellular fractions. 
The factor of $2$ in \eqref{eq:AIC} counts the fact that we estimate as primary parameters the mean and variance of each cluster. Also, in our dataset, we have $d = 11$ fractions. In sum, our proposed AIC index is a penalized model goodness of fit, which attempts to balance ("penalize") the in-sample goodness of fit with an adjustment factor that penalizes more complex models. Note, our choice in \eqref{eq:AIC} is a relatively conservative criterion that counts the clusters and fraction levels as separate and independent. 



\begin{figure}[H]
    \centering
    \includegraphics[width=\linewidth]{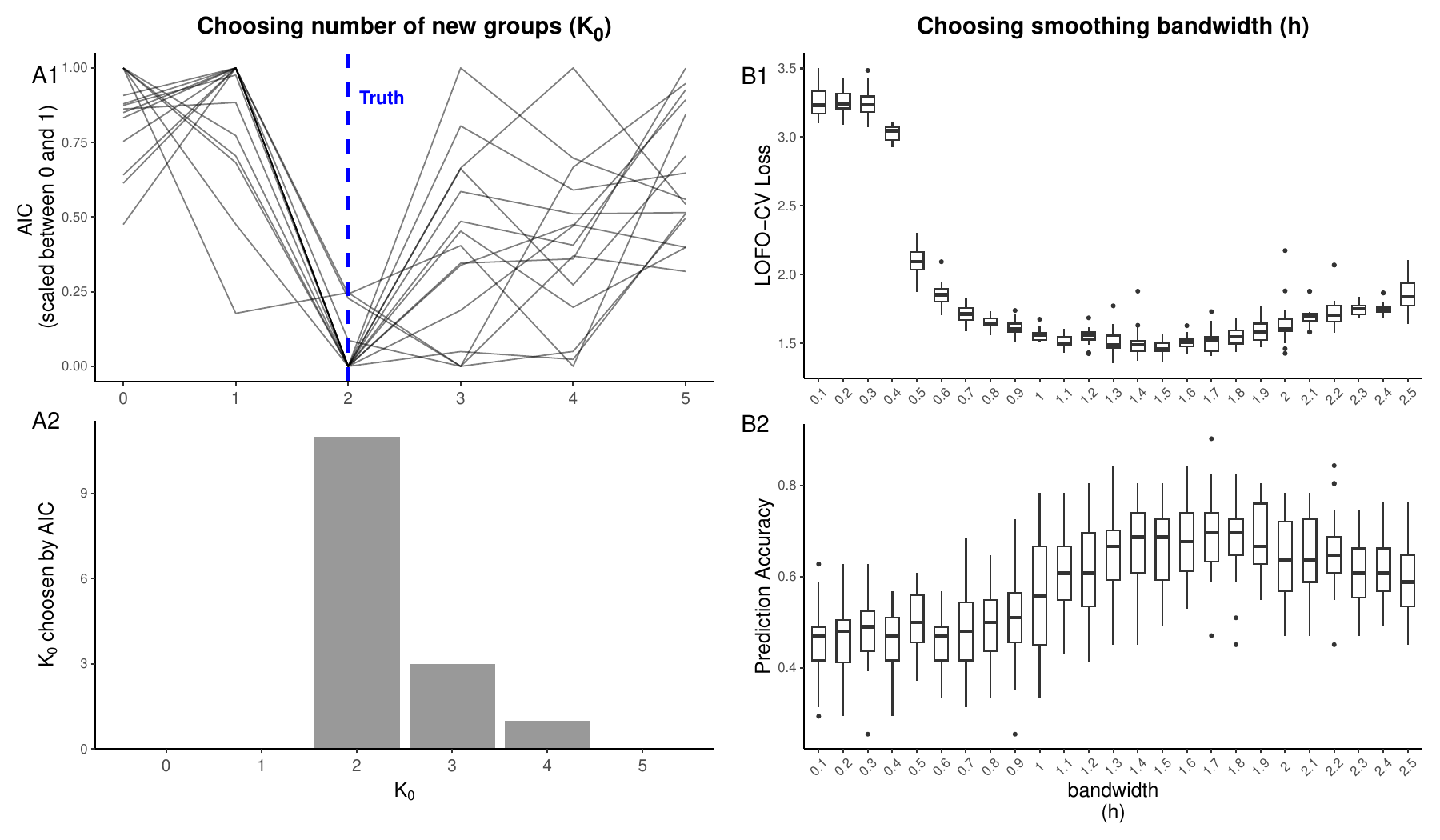}
    \caption{\textbf{Hyper-parameter selection.} 
   \textbf{(Panel A1)} AIC curves from an experiment (described in Section~\ref{sec:selectK0}) where two known protein groups are masked, then AIC is used to select $K_0$. Each curve has been normalized to be between $0$ and $1$ for ease of visualization. \textbf{(Panel A2)} The relative frequency of the minimizing value of $K_0$ in each of the 15 curves from Panel A1. The correct value $K_0=2$ is chosen 11 out of 15 times.
   \textbf{(Panel B1)} Leave-one-fraction-out cross-validation (LOFO-CV) scores over a grid of bandwidth values $h$ which controls the smoothness of the group protein means in the \texttt{FSPmix} model. Each boxplot shows the result of leaving out one of the 11 fractions out at a time. In the next panel, we compare the result of this to an alternative experiment.
   \textbf{(Panel B2)} Hold-out prediction accuracy (from a 60/40 train-test split) of group memberships. Each boxplot shows the test-set classification accuracy over many replicates. The accuracy peaks in approximately the same value of $h$ where the LOFO–CV score attains its minimum. }
    \label{fig:validating_choosing_hyper}
\end{figure}

Next, we performed a simulation study on our Thaps2024  dataset to verify the ability of our proposed AIC criterion \eqref{eq:AIC} to correctly recover the true number of undiscovered protein groups. We withheld the labeled proteins from two groups and estimated a model using data from the remaining labeled proteins from six groups. The upper-left panel of Fig \ref{fig:validating_choosing_hyper} overlays the resulting AIC profiles across ${6 \choose 2}$ replicates for models using $K_0$ values of zero through five. In the vast majority of replicates, the profile exhibits a sharp minimum at $K_0=2$, precisely matching the number of groups that were withheld. The lower-left histogram aggregates these minima: $K_0=2$ is by far the most frequent value chosen "(11 out of 15), while $K_0>2$ is chosen much more infrequently and $K_0<2$ is never chosen. These findings indicate that selecting $K_0$ using AIC is accurate and seldom results in an overly complex model with superfluous groups.





\section{Numerical Experiments}
\label{sec:simulations}
In this section, we present two numerical experiments designed to assess the practical performance of the proposed semi‑supervised mixture model on the subcellular proteomics data sets introduced in section \ref{sec:data}. 
We conducted two complementary experiments. In Section~\ref{sec:pred_exist}, we evaluate the model's ability to assign proteins to known subcellular protein groups, benchmarking its predictive accuracy against established supervised and semi‑supervised methods. In Section~\ref{sec:pred_new}, we design an experiment to probe the model’s capacity to uncover previously unannotated groups by manually withholding labels from a subset of known groups and comparing the inferred new clusters with withheld groups. We start by assessing how well the model classifies proteins into known subcellular groups before showing its capacity for new group discovery.


\subsection{Predicting protein group membership}
\label{sec:pred_exist}

In order to evaluate our model's ability to assign proteins to known subcellular niches, we performed a numerical experiment from the real data. As a first step, we tuned $h$ once with the entire Thaps2024 dataset. We then repeatedly took a random group-stratified 40\% of the labeled data and masked these labels. Then, we fit a FSPmix model with $K_0 = 0$ on this partially labeled data (with only 60\% of the original labels) and made soft label predictions (using \eqref{eq:coinflip}) on the masked proteins. Lastly, we revealed the ground-truth labels that we had masked, and computed the proportion of correctly predicted labels out of the total number of masked proteins.


We compared our approach to several other methods -- k-nearest neighbours (KNN), support vector machines (SVM) from the pRoloc \cite{gatto2014mass} R package, and Label Propagation 
\citep{Zhou2004Learning} 
from the LAMBDA\_SSL \cite{jia2024lamda} package in Python. 
(See Appendix~\ref{sec:cross-validation-details} for more details.)
Also, for a thorough comparison, we used not only our Thaps2024 dataset but also five other published spatial proteomics datasets described in the pRolocdata R package \cite{gatto2014mass}. Briefly, \textit{E14TG2aR} \cite{christoforou2016draft} was generated from mouse embryonic stem cells; \textit{hirst2018} \cite{hirst2018role} was generated from human HeLa cells; \textit{lopitdcU2OS2018} \cite{thul2017subcellular} was generated from human U2OS cells; \textit{moloneyTbBSF} \cite{breckels2016learning} was generated from \textit{Trypanosoma sp}. cells, and \textit{yeast2018} \cite{nightingale2019subcellular} was generated from \textit{Saccharomyces cerevisiae} cells. These different datasets were obtained using various subcellular fractionation and protein quantification approaches. Importantly, Thaps2024 and E14TG2aR are ``low-replicate", i.e. containing data from a single experimental replicate, while the other datasets had ``high replication'', i.e. two or more experimental replicates. By distinguishing between low- and high-replicate datasets, we are able to better evaluate the performance of our statistical method in both data-limited and data-rich regimes.

\begin{figure}[H]
    \centering
    \includegraphics[width=\linewidth]{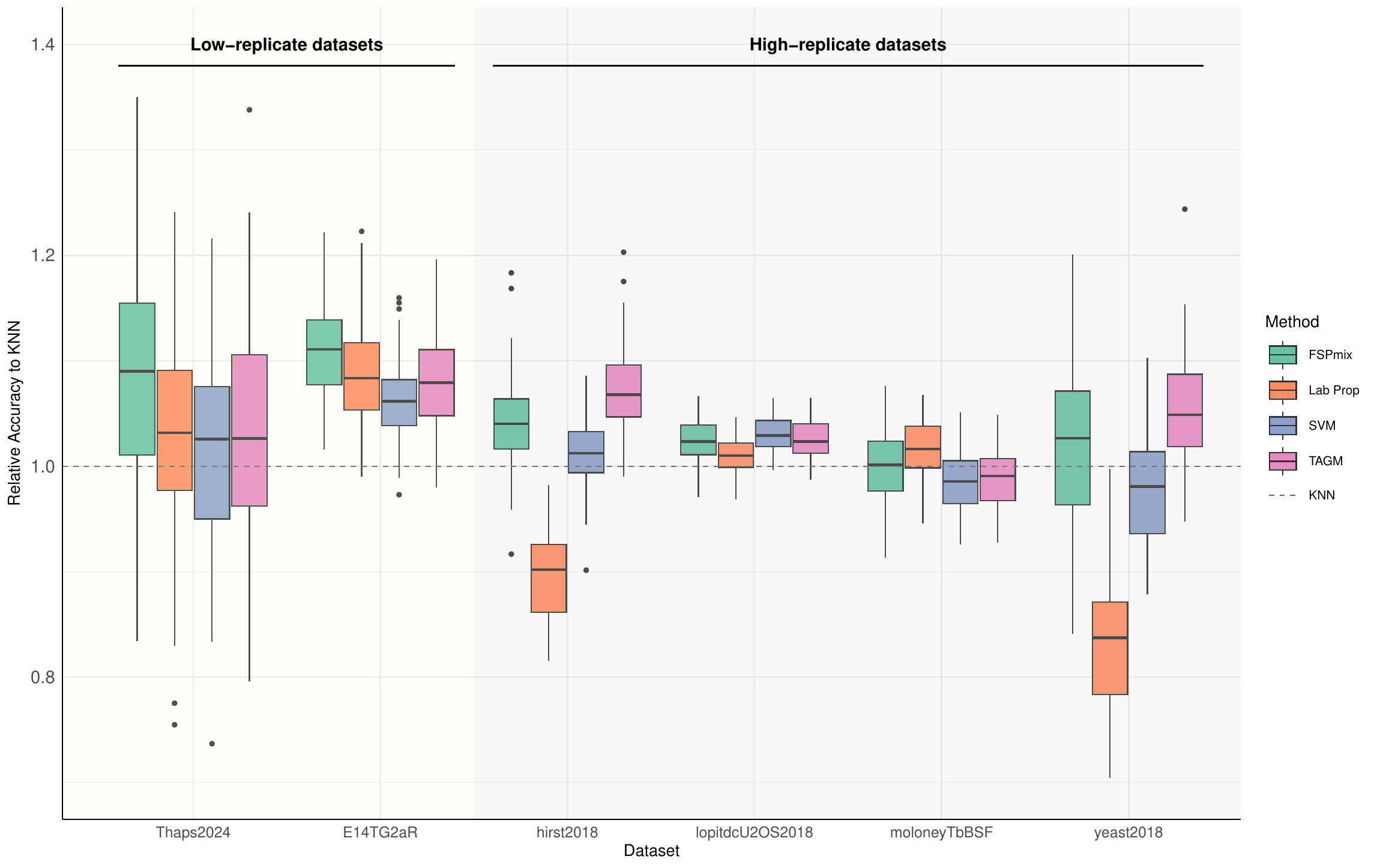}
    \caption{\textbf{Four semi-supervised classifiers applied to six spatial-proteomics datasets.}     Each set of boxplots shows the relative group membership prediction accuracy of four models (FSPmix, label propagation (Lab Prop), support vector machines (SVM), and TAGM) from $100$ simulations (described more in detail in Section~\ref{sec:pred_new}. Here, relative accuracy is the ratio of the accuracy of each method to that achieved by K-nearest-neighbor (KNN), which serves as the baseline (dashed line at 1.0). Values above the dashed line indicate that the method outperforms KNN. FSPmix outperforms other methods especially in the ``low-replicate" datasets containing only a single biological replicate.  (Absolute accuracy is shown in appendix Fig~\ref{fig:pred_acc}.).}
    \label{fig:acc_comparison}
\end{figure}

Fig~\ref{fig:acc_comparison} shows the test-set accuracy of FSPmix, SVM, and TAGM, each normalized by the accuracy of the KNN classifier in the same train–test split across all six benchmark data sets.
Across all six datasets, FSPmix attains parity or a clear improvement relative to the baseline, and it is never dominated by other methods. The gain is most pronounced on E14TG2aR and Thaps2024, the two smallest data sets in the panel, where FSPmix’s median relative accuracy reaches approximately  $1.1$.  In the outlier-rich hirst2018 dataset, the performance of FSPmix is somewhat poor compared to that on other datasets, but it remains competitive overall; this indicates that the FSPmix model's Gaussian mixtures are somewhat sensitive to outliers or heavy-tailed noise. 
On the remaining larger data sets (lopitdcU2OS2018, moloneyTbBSF, yeast2018), FSPmix matches or modestly exceeds the baseline and tracks closely with TAGM. Collectively, these results suggest that FSPmix is particularly advantageous in a low signal-to-noise scenario, while highlighting a potential avenue for improvement in the presence of pronounced outliers.

\subsection{Comparing performance of new protein group discovery}
\label{sec:pred_new}

Next, in order to assess the model’s capacity for new protein group discovery, we conducted an experiment in which we simulated discovery of unknown groups. Starting with the labeled data in the Thaps2024 dataset containing six known groups, we randomly masked two groups, treating all proteins in these groups as unlabeled, and then fitted the FSPmix model using labels only from the four remaining groups. We selected the optimal $K_0$ value using AIC which was introduced in \ref{sec:selectK0}.

After training, we cross-tabulated the predicted group labels of the masked proteins against their ground-truth group labels and applied the Hungarian algorithm (maximum-weight bipartite matching) \cite{kuhn1955hungarian} to 
find the two discovered groups most aligned with the two masked groups.
Then, we calculated the proportion of proteins whose group membership was correctly estimated, out of all proteins in the two masked groups. Such a classification accuracy aims to quantify the model's ability to identify entirely unseen groups.




We additionally used three other methods for protein group discovery to compare with FSPmix -- a regular GMM assuming the covariance structure specified in \ref{eq:sigma}, Phenodisco \cite{breckels2013effect}, and TAGM\_novel \citep{crook2020semi-disco}. For the latter two methods, we used their default procedures for selecting the number of groups $K$, while for the regular GMM we used the same number of groups chosen for FSPmix using AIC.

\begin{figure}[H]
    \centering
    \includegraphics[width=\linewidth]{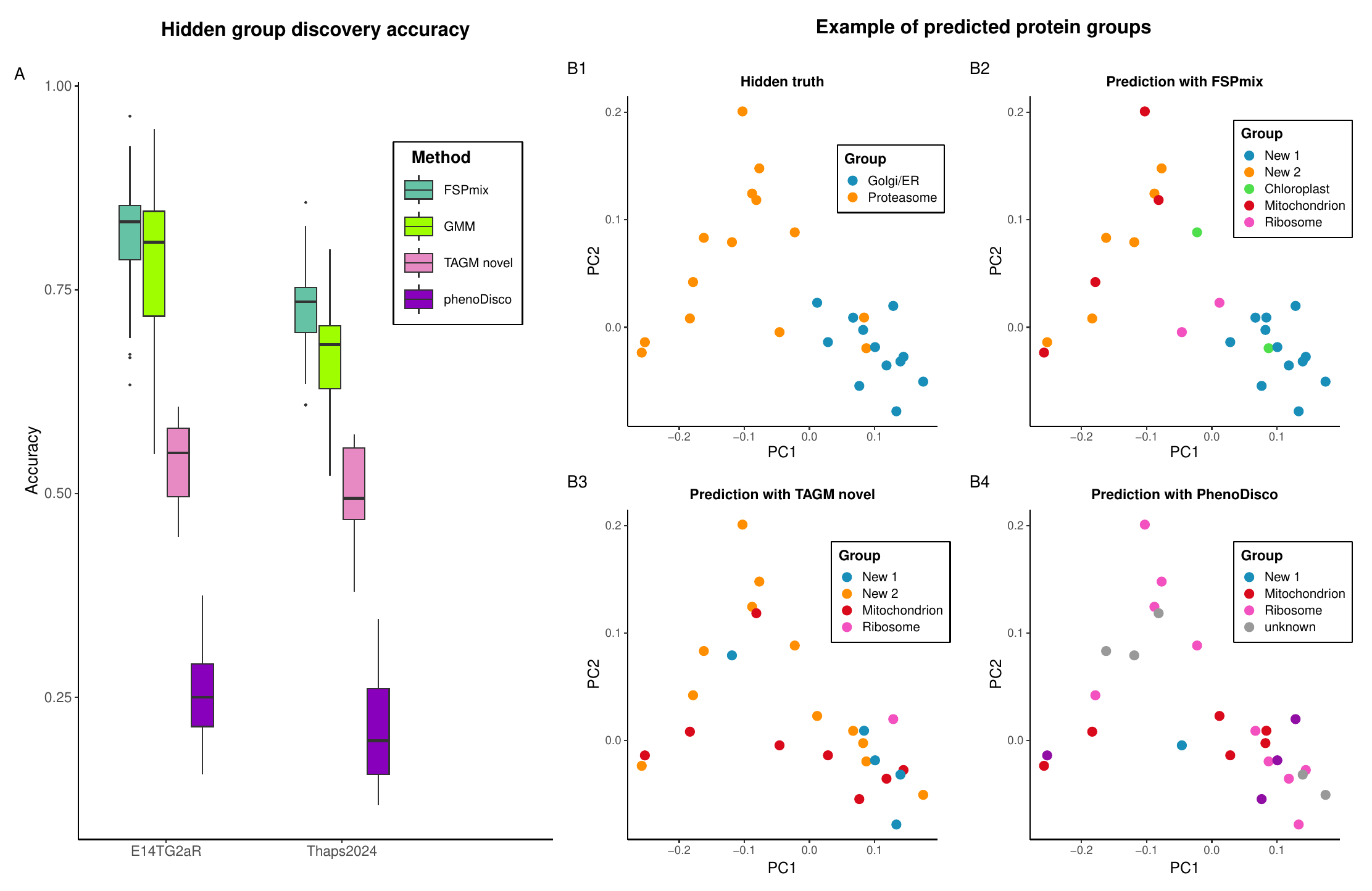}
    \caption{\textbf{New protein group discovery.}  In a simulation, we masked the labels of two protein groups, and used four models to predict new group memberships on these masked groups. \textbf{(Panel A)} The boxplots summarize the classification accuracy (among the masked proteins) across 30 simulations. \textbf{(Panel B1-B4)} One representative simulation is visualized. 
    \textbf{B1} shows, in 2d PCA space, proteins from two groups (Golgi/ER and Proteasome) whose labels we mask. \textbf{Panels B2-B4} show as colors the predicted labels of these proteins from three models (note: the colors were algorithmically reordered to best match the true group labels.) Only FSPmix predicts these masked proteins to belong to two distinct new clusters.}
    \label{fig:New group discovery V2}
\end{figure}

Fig \ref{fig:New group discovery V2} summarizes the results of this experiment. The boxplots in Fig~\ref{fig:New group discovery V2} (panel A) show that FSPmix outperforms PhenoDisco on both datasets in terms of the aforementioned accuracy metric. Next, Panel B visualizes one representative simulated trial to illustrate how various methods make predictions on real data. Fig \ref{fig:New group discovery V2} panel B1 shows the masked true labels of two protein groups Golgi/ER and Proteasome, and panel B2 illustrates FSPmix predictions clearly recovering these two groups as distinct new groups (New2 and New1 respectively). On the other hand, Fig \ref{fig:New group discovery V2}  panels B3 and B3  show that both Phenodisco and TAGM-novel do not identify those two groups as separate, and make erroneous predictions. 

We briefly discuss the implications of this. The weakest performance comes from the PhenoDisco model, which first applies PCA for dimensionality reduction and then fits a Gaussian mixture in the reduced 2-dimensional space. Although PhenoDisco may perform well on datasets with well-separated protein signatures, its reliance on PCA as a first step makes it less competitive when applied to heavily overlapping protein signatures. Next, while FSPmix and GMM both assume a diagonal $d \times d$ covariance matrix for each cluster's protein signatures, TAGM\_novel requires estimation of the full variance-covariance matrix. For low S/N data, such a high-dimensional covariance structure is estimated poorly because the number of parameters can be large compared to the sample size. We interpret this as the main reason why FSPmix and GMM both outperform TAGM\_novel in Thaps2024 dataset. Lastly, FSPmix outperforms the regular GMM because it incorporates a smooth mean function across size fractions. By taking information from neighboring fractions toward each other, this smoothness improves the robustness of the estimates when data are limited.

\section{Case study with Thaps2024 dataset}
\label{sec:real-data}

In this section, we applied our semi‑supervised functional mixture model FSPmix to the Thaps2024 diatom subcellular proteomics dataset. We began by selecting the kernel bandwidth via leave-one-fraction-out cross validation (LOFO-CV) which was introduced in Section~\ref{sec:LOFO}. Then, using the selected bandwidth $h$, we estimated two FSPmix models; the first is one with a fixed number of groups $K=6$, equal to the number of manually annotated subcellular niches (so that $K_0=0$). We call this the null model. The second model we estimated used $K=10$ which was chosen by AIC as described in \ref{sec:selectK0}. We will refer to this as the selected model. The null model suppresses new group discovery, and assumes that the marker proteins have revealed all existing groups, while the selected model is allowed the flexibility to discover four new groups of proteins whose profiles are coherent with each other.
Thus, the comparison between these two models allows us to assess the extent to which the FSPmix model identifies biologically meaningful clusters of proteins beyond the initial manual annotations.


To carefully examine the biological implications of our proposed FSPmix model on the Thaps2024 dataset, we compare the two models (null and selected) in several ways:
\begin{itemize}
\item[(i)] We visualize and study group assignments for every unlabeled protein in the Thaps2024 dataset (Section~\ref{sec:real1}).
\item[(ii)] We produce and analyze high‑probability bands that summarize uncertainty in the original expression space (Section~\ref{sec:real2})
\item[(iii)] We examine a few specific proteins whose predicted groups (or ``localisations") differ between the two models, assessing whether the selected model yields biologically more plausible group reassignments to those proteins. (Section~\ref{sec:real3}).
\end{itemize}


\subsection{Membership prediction for unlabeled proteins} 
\label{sec:real1}

\begin{figure}[H]
    \centering
    \includegraphics[width=0.9\linewidth]{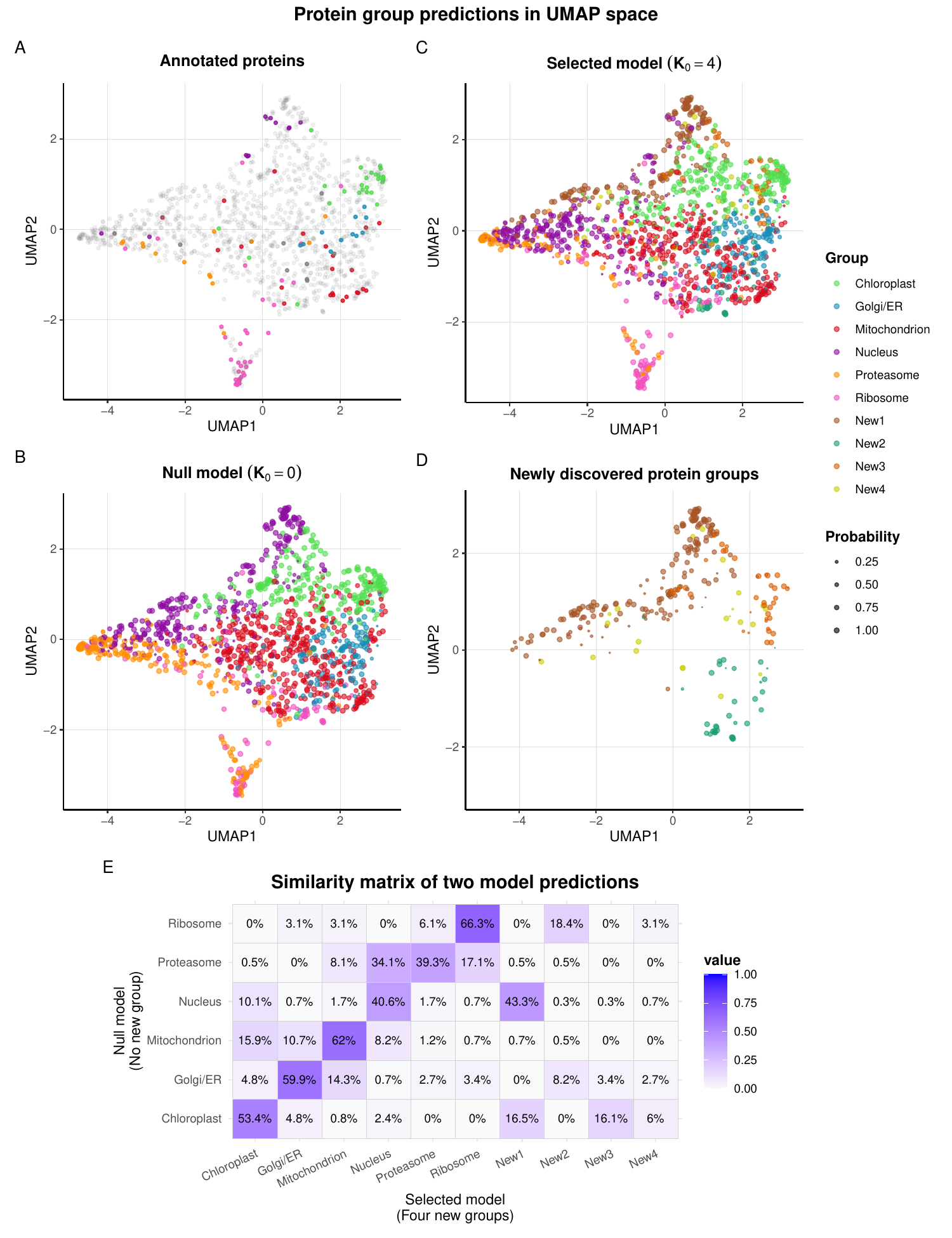}
    
    \caption{\textbf{UMAP visualization of predicted protein groups.} \textbf{(Panel A-D)} In each panel, a point represents a protein, and point size encodes the predicted group membership probability. Marker proteins are shown as colored points, and unlabeled proteins are shown in grey. Panel B shows protein group predictions from the ``null'' model ($K_0=0$) that suppresses new protein group discoveries.  Panel C shows predicted protein groups from the ``selected'' model ($K_0=4$) which allows the discovery of four new protein groups. 
    Panel D shows only the proteins predicted to belong to newly discovered groups by the selected model. \textbf{(Panel E)} A heatmap shows the cross-tabulation of predicted protein memberships between the two models, normalized so that each row sums to one. Off-diagonal elements indicate how many proteins that were initially assigned to a single group (when $K_0=0$) were re-allocated to a newly discovered group, highlighting biologically interpretable group reallocations identified by the FSPmix model with $K_0=4$.}
    \label{fig:umap-summary}
\end{figure}

Fig \ref{fig:umap-summary} contrasts the group predictions of every proteins in Thaps2024 dataset made by the null ($K=6$) and the selected ($K=10$) model. In all panels, the data have been projected onto a two-dimensional UMAP space. Panel A of Fig~\ref{fig:umap-summary} shows that, even for the labeled proteins, there is substantial overlap among different groups, which highlights the low signal-to-noise nature of the Thaps2024 dataset. Next, panels B and C show the predicted results for null model and selected model respectively. We can see that most of the labeled groups form compact ``islands" that center around their labeled proteins.  Also, expanding the mixture from six to ten components leaves these core islands intact yet isolates four coherent new clusters that appear along the peripheral, low‑density rim of the manifold (panel D of Fig~\ref{fig:umap-summary}), precisely where ambiguous or "boundary" proteins concentrate -- regions typically inhabited by unlabeled proteins that do not fit cleanly into established niches. We study this in more detail, next.

The heatmap in panel E of Fig \ref{fig:umap-summary} visualizes a row‑normalized similarity matrix of the two models' group predictions. The large values on the diagonal of this matrix confirm that most proteins retain their original localisation even by the more complex, selected model that aims to discover four more groups. Also, the off‑diagonal elements highlight biologically interpretable \textit{reallocations} of the selected model -- a subset of proteasomal proteins splits between the nuclear and ribosomal groups; several boundary Mitochondrial proteins co‑localise with Golgi/ER (and, to a lesser degree, chloroplast); and a sizeable portion of nuclear proteins is assigned into the new group New1. Furthermore, two new groups (New2 and New3) derive chiefly from the outer fringes of the Ribosome and Chloroplast groups, in UMAP space. Appendix figures \ref{fig:by-null} and \ref{fig:by-selected} show a more detailed group-by-group comparison, and Appendix Fig~\ref{fig:alluvial} further compares the protein-by-protein predictions of the null model and selected model in detail. Together, these patterns indicate that the selected model preserves established subcellular niches while uncovering finer‑grained, biologically plausible new niches.  

\begin{figure}
    \centering
    \includegraphics[width=0.8\linewidth]{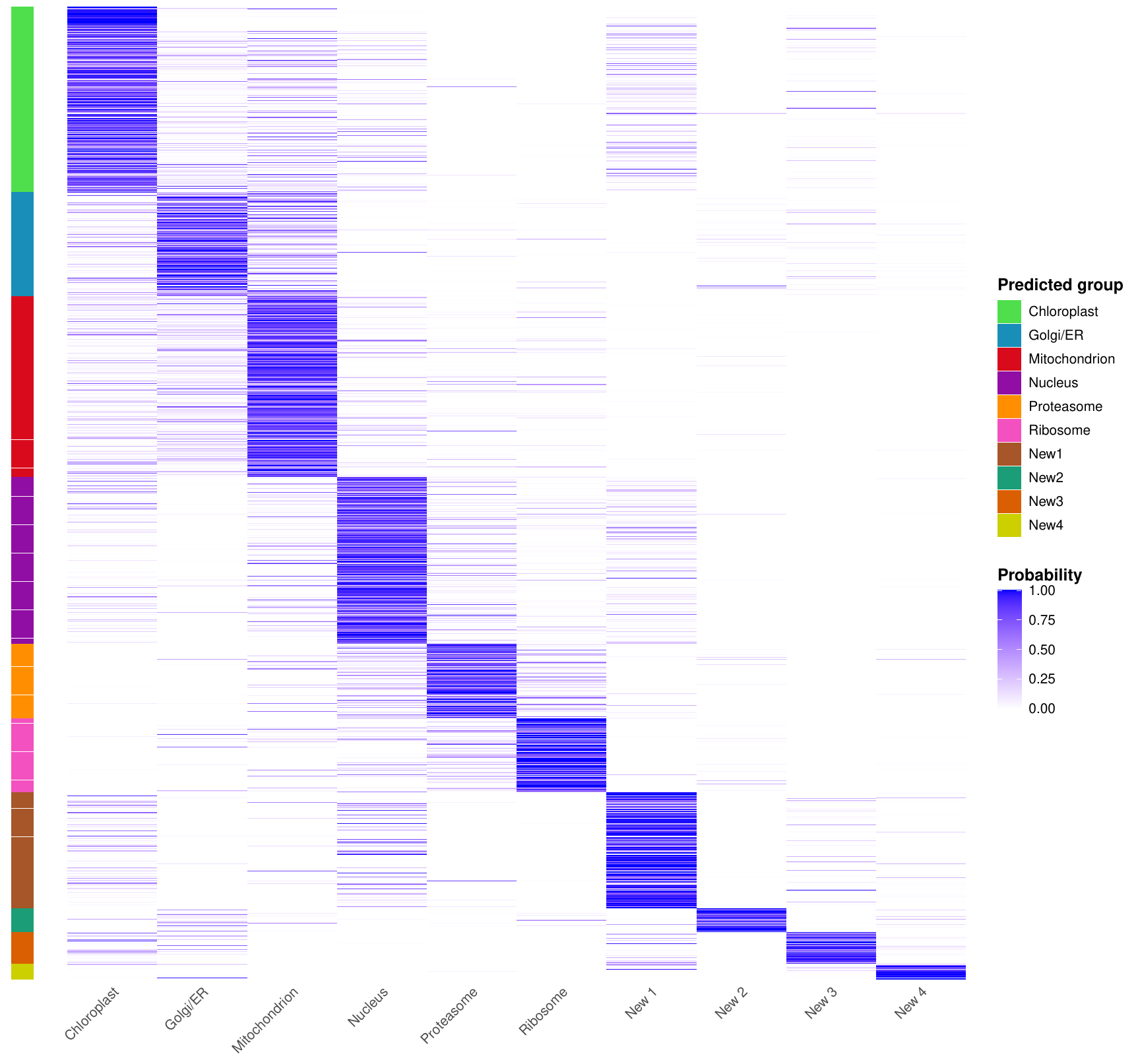}
    \caption{\textbf{Matrix representation of estimated group probabilities of all proteins.} Each matrix row -- a thin horizontal line -- represents the $K$ estimated responsibilities (group membership probability) of one protein. The colored bars on the left-hand side show the predicted group of each protein. 
   The rows of this matrix are ordered by predicted group assignment, resulting in prominent diagonal blocks. These blocks are not uniformly shaded deep blue because the final predicted groups are based on a random multinomial draw from the estimated probabilities, as described in Section~\ref{sec:prob-estimation}.
   Thus, off-diagonal regions represent the ambiguity between predicted groups and estimated groups. For a given predicted group, deeper blue shading in other off-diagonal columns visually signifies a lower confidence in that prediction and indicates that the groups in question are less well-separated in protein signature space.}
    \label{fig:responsibility}
\end{figure}

Fig~\ref{fig:responsibility} further visualizes the statistical uncertainty in the protein memberships predictions made by the selected model (10 groups). Every row represents a protein, and each column of a given row shows the estimated membership probabilities $ \{\hat\gamma_{ik}\}_k$. The rows are grouped by their soft membership predictions made using \eqref{eq:coinflip}.  We can see that the majority of proteins fall along the diagonal area, indicating that the model confidently predicts them to be of a single subcellular group. However, some proteins (rows) have non-negligible probabilities in the off-diagonal (notably those predicted to be Chloroplast, Mitochondrion, and Golgi/ER). For these proteins, the soft membership prediction is less confident due to the significant overlap in the protein curves and marker proteins -- leading to estimated membership probabilities are less concentrated solely on one group. These observations highlight FSPmix's ability to effectively quantify uncertainty in protein memberships.

\subsection{Prediction bands for protein groups} 
\label{sec:real2}

\begin{figure}
    \centering
    \includegraphics[width=\linewidth]{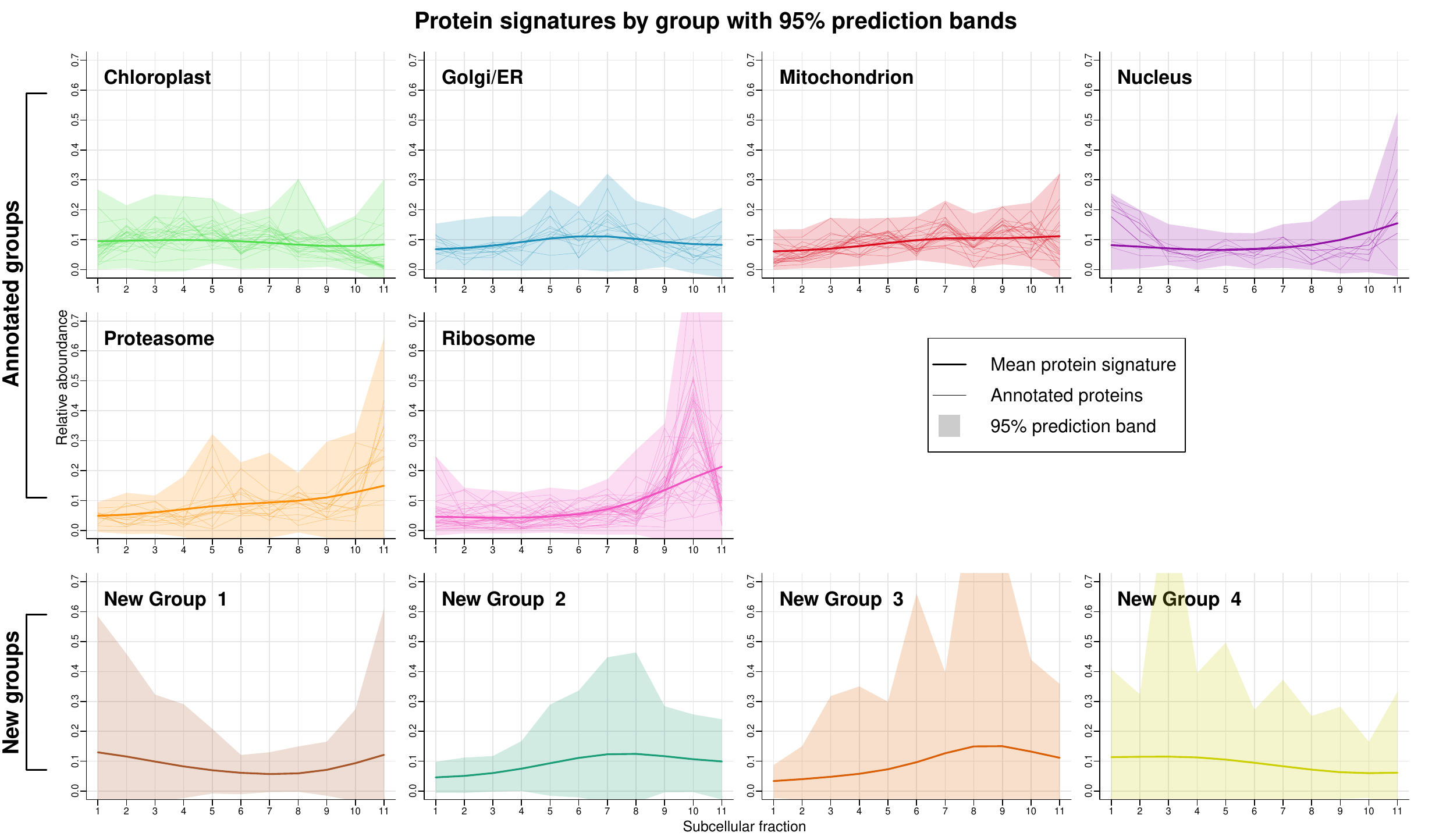}
    \caption{\textbf{Simultaneous 95\% prediction bands for each protein group.} 
    Each panel depicts shows prediction bands for fraction‐wise abundance profiles of a single protein group. Groups with annotated marker proteins are shown the first two rows, and new discovered clusters are shown in the bottom row. In each panel, a thick line shows the estimated smooth group mean profile, and the thin lines show the labeled marker proteins. The shaded region shows the cluster-conditional simultaneous 95\% prediction bands introduced in section~\ref{sec:high-prob-region}. 
    The prediction bands successfully encompass the vast majority of marker proteins without being unnecessarily wide, indicating both high empirical coverage and statistical power.
    All predicted unlabeled protein signatures are fully plotted in Fig~\ref{fig:original}. 
    (The y-axis is top-truncated slightly for ease of visualization.)} \label{fig:hpb}
\end{figure}

We next construct the 95\% prediction bands obtained with the procedure described in Section~\ref{sec:high-prob-region} for the selected FSPmix model (with $K=10$). The results are shown in Fig~\ref{fig:hpb}. For the six labeled groups (the top two rows of panels of Fig~\ref{fig:hpb}) the bands are characteristically narrow and closely track well‑known subcellular signatures. Specifically, chloroplast and mitochondrion protein signatures remain almost flat across fractions, whereas ribosome and proteasome protein signatures rise sharply towards the latter fractions. The simpler null model -- which disallowed the discovery of new groups -- produced noticeably wider prediction bands, especially for Chloroplast and Ribosome panels (Fig~\ref{fig:no-extra}). Importantly, when using the selected model in Fig~\ref{fig:hpb}, almost every manually annotated protein lies entirely within the ribbon of its corresponding subcellular niche. This, from an empirical perspective, confirms that the nominal 95\% coverage is achieved and that the clustering is consistent with established biological knowledge.

Next, we see that the four newly discovered protein groups (bottom row of Fig~\ref{fig:hpb}) have significantly broader prediction bands than the groups with annotated marker proteins (top row). This greater uncertainty of new groups is expected, as these new discovered protein groups were estimated without marker proteins. Despite this larger variability, each new protein group retains a recognizable and distinct mean profile. For example, New Group 1 shows a slight decline in middle fractions, whereas New Group 2 peaks slightly in these fractions. The visualization of these new group mean curves, together with their clear separation from labeled groups, suggests that the model is isolating biologically meaningful and previously unrecognized subcellular groups.

\subsection{Subcellular protein localization examples}
\label{sec:real3}

\begin{figure}
    \centering
  \includegraphics[width=\linewidth]{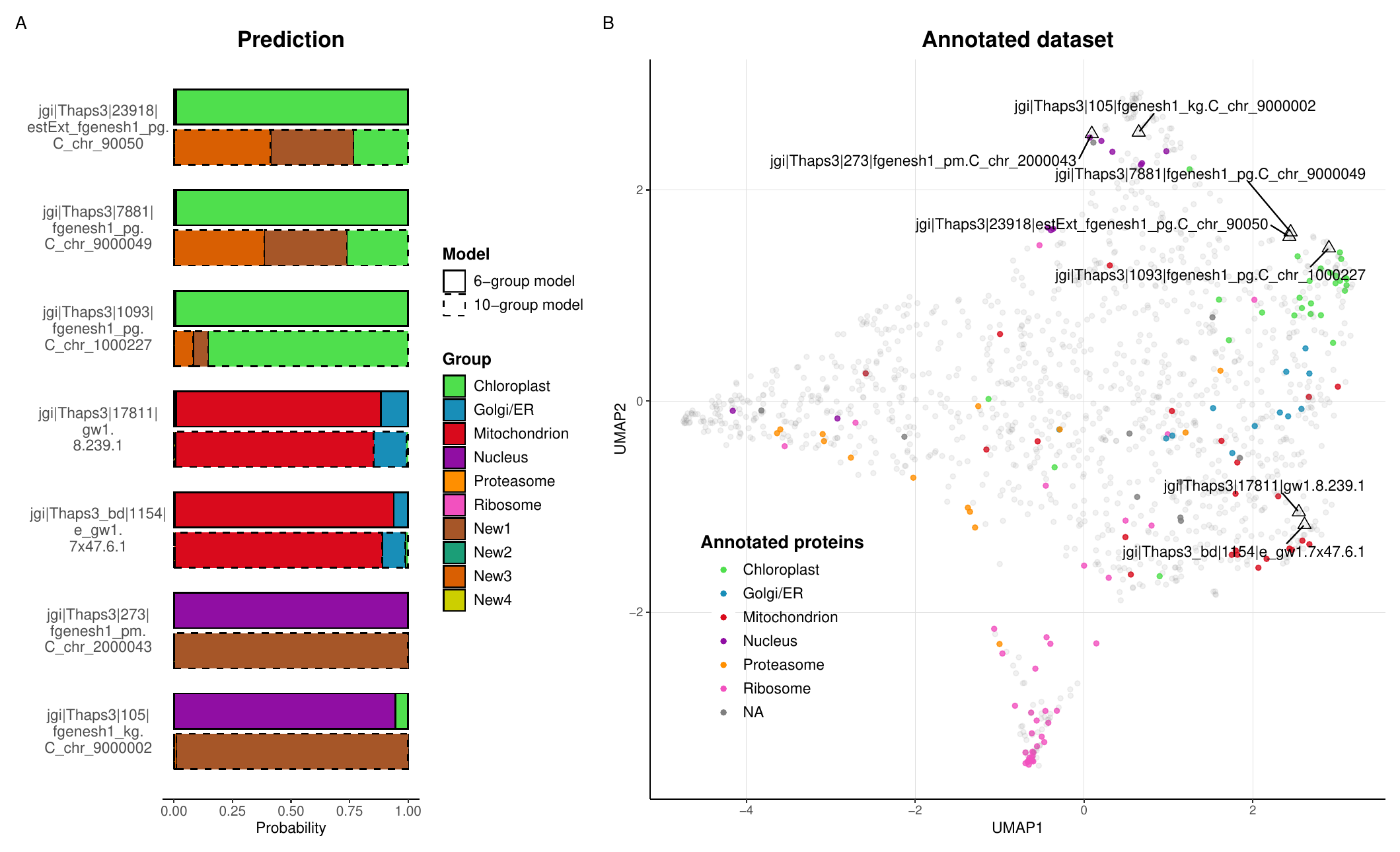}
    \caption{\textbf{Subcellular localization predictions for selected proteins.}  This figure examines the importance of new group discovery allowed by FSPmix, focusing on seven well-studied proteins.
    \textbf{(Panel A)} Two barplots are shown for each of the seven proteins, one showing the estimated group membership probabilities (responsibilities) from a six-group FSPmix model ($K=6, K_0=0$) that suppresses new discovery, and one from a ten-group model that aims to find four new groups ($K=10, K_0=4$). 
    The predicted probabilities are frequently very different between the two models, which signifies the importance of allowing the FSPmix framework the flexibility to discovery new groups.
    \textbf{(Panel B)} Two dimensional UMAP visualization that shows the relative position of these seven proteins, overlayed with the marker proteins (as colored points), and all other un-annotated proteins (as grey points). This panel provides visual context for the probabilitistic predictions shown in Panel A.}
    \label{fig:speci_pred}
\end{figure}

To evaluate whether our clustering method assigns proteins to their correct subcellular niches, we selected seven proteins (Fig \ref{fig:speci_pred}) that were not included in the training dataset and predicted their localization using our 10-group and 6-group models.  

Proteins 23918, 7881, and 1093 were strongly predicted to localize to the chloroplast using both our 10- and 6- group model (Fig~\ref{fig:speci_pred}) - taking into consideration that groups New1 and New3 are likely to be clusters belonging to the chloroplast (Fig~\ref{fig:umap-summary}). Of these three proteins, 7881 and 23918 were until recently classified as having unknown functions, but recent imaging and knockout experiments have localized them to the chloroplast as components of the PyShell structure \cite{shimakawa2024diatom}. Protein 1093 is functionally annotated as ``magnesium transporter", and may play a role in regulating magnesium homeostatis in the chloroplast for chlorophyll biosynthesis  \cite{zhang2022two}.  These examples illustrate the utility of our approach in uncovering new protein functions based on subcellular localization. Notably, the PyShell proteins are accurately placed within the chloroplast in our data, despite their functional roles being only recently discovered.

Proteins 17811 and 1154, both functionally annotated as members of the Triose-Phosphate Transporter (TPT) family, were predicted by our models to localize to the mitochondrion. Interestingly, TPT proteins are typically known to localize to the chloroplast \cite{moog2015localization}. This discrepancy may reflect methodological limitations—such as challenges in cell lysis or organelle separation. Alternatively, it could reflect the close physical associations between mitochondria and chloroplasts in diatoms, including membrane contact sites  (\cite{prihoda2012chloroplast}, \cite{flori2017plastid}), or suggest that these proteins may play novel functions within mitochondria in diatoms. 

Proteins 273 and 105, both annotated as translation-initiation factors, were predicted to localize to the nucleus in both the 10-group and 6-group models. Although translation is generally initiated in the cytoplasm, studies in human cells have shown that these proteins localize to the nucleus during normal growth, and relocate to the cytoplasm under stress, where they may perform other functions \cite{grove2023increased}. Because our Thaps2024 dataset was generated from healthy diatom cells in mid-exponential growth, we expect localization to the nuclear group, which our model corroborates. These findings additionally underscore the potential value of investigating subcellular protein localizations under varying environmental conditions to elucidate protein functions \citep{Crook2022}.

\section{Conclusions}
\label{sec:discussion}
This work introduces a semi‑supervised functional clustering framework that combines smoothed mean trajectories with a finite‑mixture likelihood to analyze subcellular spatial proteomics data in which both observations and manual annotations are scarce. By borrowing statistical strength across fractions and across proteins both labeled and unlabeled data, the method delivers three outputs that are critical to cell‑biology workflows: (i) model based probabilistic assignments of every protein to its predicted subcellular niche; (ii) interpretable mean signatures for each subcellular niche and cluster‑conditional prediction bands that quantify uncertainty; and (iii) principled discovery of previously uncharacterised protein groups.

Comprehensive benchmarks highlight the method’s strengths: In low signal-to-noise datasets where classical supervised classifiers suffer from over‑fitting and purely unsupervised algorithms ignore a wealth of partial labels, our approach achieves state-of-the-art prediction accuracy. On label-rich or high-replication datasets, it remains competitive with leading baselines while uniquely offering uncertainty quantification and novel cluster detection. In synthetic evaluations designed to test novelty, the model recovers hidden clusters more reliably than clustering-based pure unsupervised workflows.

A case study on the Thaps2024 spatial proteomics dataset demonstrates translational impact. When applied to previously unlabeled proteins, our model not only accurately predicted their localization to subcellular niches consistent with known functional annotations, but also revealed subtler patterns of localization that were not apparent from functional annotation alone. 

Limitations suggest fruitful directions for future work. First, this model inherits the sensitivity of mixture model to heavy‑tailed noise. Second, when protein signatures exhibit little smoothness, or when replication is so high that protein trajectories are very easy to be seperate from each other, our gains over baseline mixture models are narrow. So, adaptive bandwidth selection, non-Gaussian noise for outlier‑prone fractions may alleviate these weaknesses. Also, marker protein signatures in spatial proteomic data have tighter distributions around their means than un-annotated proteins, which suggests an extension of this model that treats the statistical distributions of the two types of proteins differently. Despite these caveats, the present work establishes a strong foundation for semi‑supervised functional clustering in subcellular spatial proteomics and, more broadly, in any high‑dimensional functional data where group labels are scarce but there is smooth structure.

\section{Acknowledgements}
\label{sec:acknowledgements}
We are indebted to Dr. Ross Waller and his group, Dr. Thomas Krueger, Dr. Kathryn Lilley, and Dr. Lisa Breckels for their invaluable insights, knowledge exchange, and discussions on spatial proteomics. This work was supported by NIH R01GM135709 and the Center for Chemical Currencies on a Microbial Planet (NSF 2019589) to M.A.S; Kimberly Foundation's Hugh Morris Experiential Learning Fellowship and the Woods Hole Oceanographic Institution to L.J.J, and the Simons Collaboration on Computational Biogeochemical Modeling of Marine Ecosystems/CBIOMES (Grant ID: 1195553) to S.H. 

\newpage

\section{Appendix}

\subsection{Appendix 1 Full details of cross-validation used in Section~\ref{sec:pred_exist}}
\label{sec:cross-validation-details}
    
In the numerical experiment conducted in Section~\ref{sec:pred_exist}, we tuned various hyperparameters using a 5-fold cross-validation. Specifically, for each simulation, we first masked one-fifth of the labels, fit the four models on the data with a particular candidate hyperparameter value, and measured as a score the proportion of the masked labels that were predicted correctly -- cycling through the 5 folds and averaging this score. We chose the best hyperparameter value that minimizes this average score. 

For each model, here are the parameters we tuned:
\begin{itemize}
    \item KNN: Neighborhood size $k$.
    \item SVM: A two-dimensional grid search on Cost $c$ and Kernel width $\sigma$ (parameter \texttt{cost} and parameter \texttt{sigma} in function \texttt{pRoloc::svmClassification()}). 
    \item Label propagation: Kernel width $\gamma$ (parameter \texttt{gamma} in function \\ \texttt{LAMDA\_SSL.Algorithm.Classification.LabelPropagation()} in Python3)
    \item FSPmix: Kernel bandwidth $h$ (parameter \texttt{bandwidth} in function \texttt{FSPmix::fspmix()})
\end{itemize}

After tuning, each model was re-fit on the full training set and evaluated on the held-out set.

\subsection{Appendix 2 Supplemental figures}
\setcounter{figure}{0} 
\renewcommand{\thefigure}{S\arabic{figure}}

\begin{figure}[H]
    \centering
    \includegraphics[width=\linewidth]{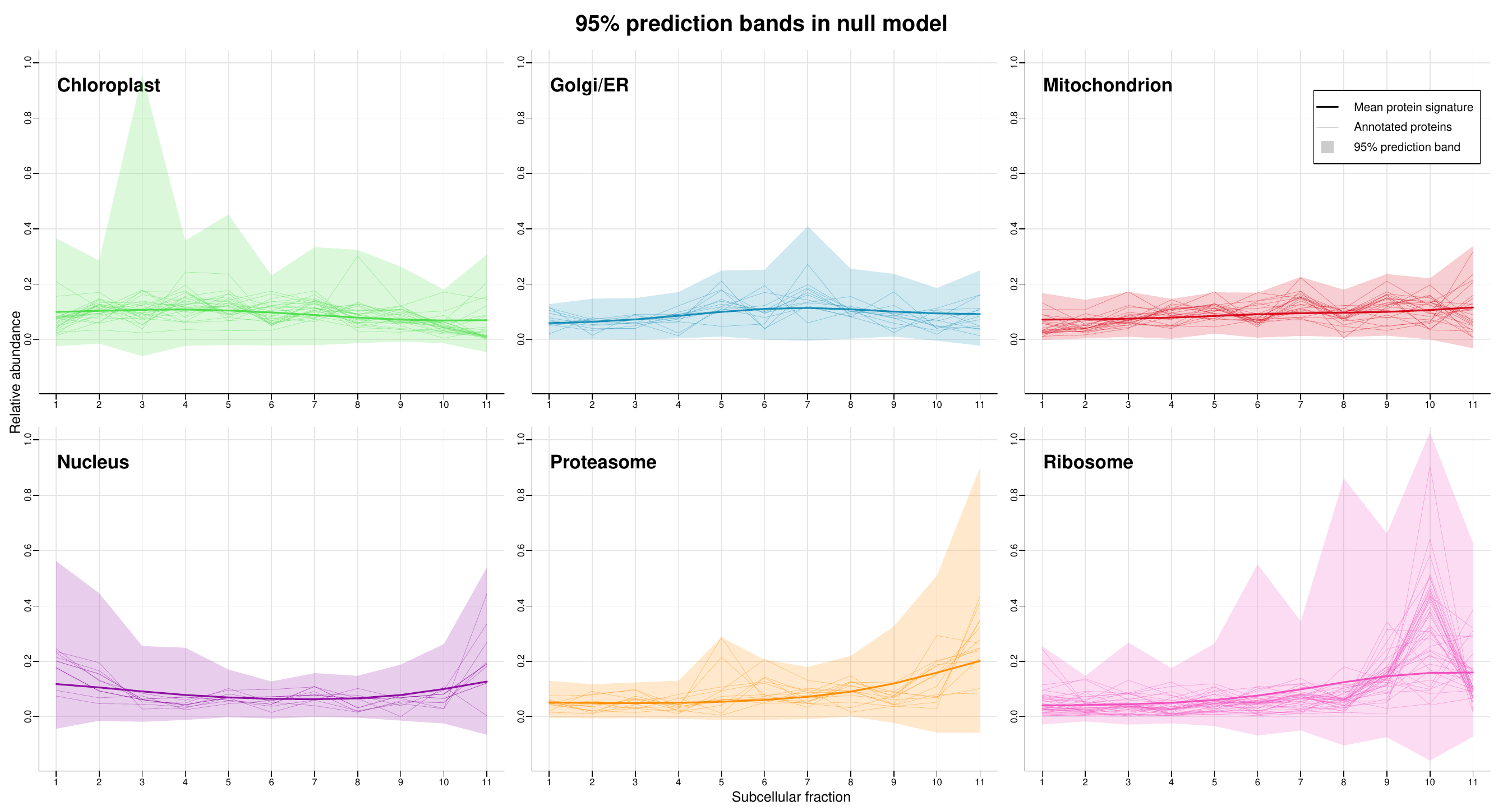}
 \caption{\textbf{95\% prediction bands for the six groups in the null model.} This figure shows prediction bands from the ``null'' model (with $K=6$, $K_0=0$) in a similar style as Fig~\ref{fig:hpb}.
 We can see that the Chloroplast prediction band in the null model is much wider than that of the selected model ($K=10$, $K_0=4$), especially in fraction 3 and 5. A similar observation can be made in Ribosome panel's fraction 6 through and 9. }
 \label{fig:no-extra}
\end{figure}

\begin{figure}[H]
    \centering
    \includegraphics[width=.8\linewidth]{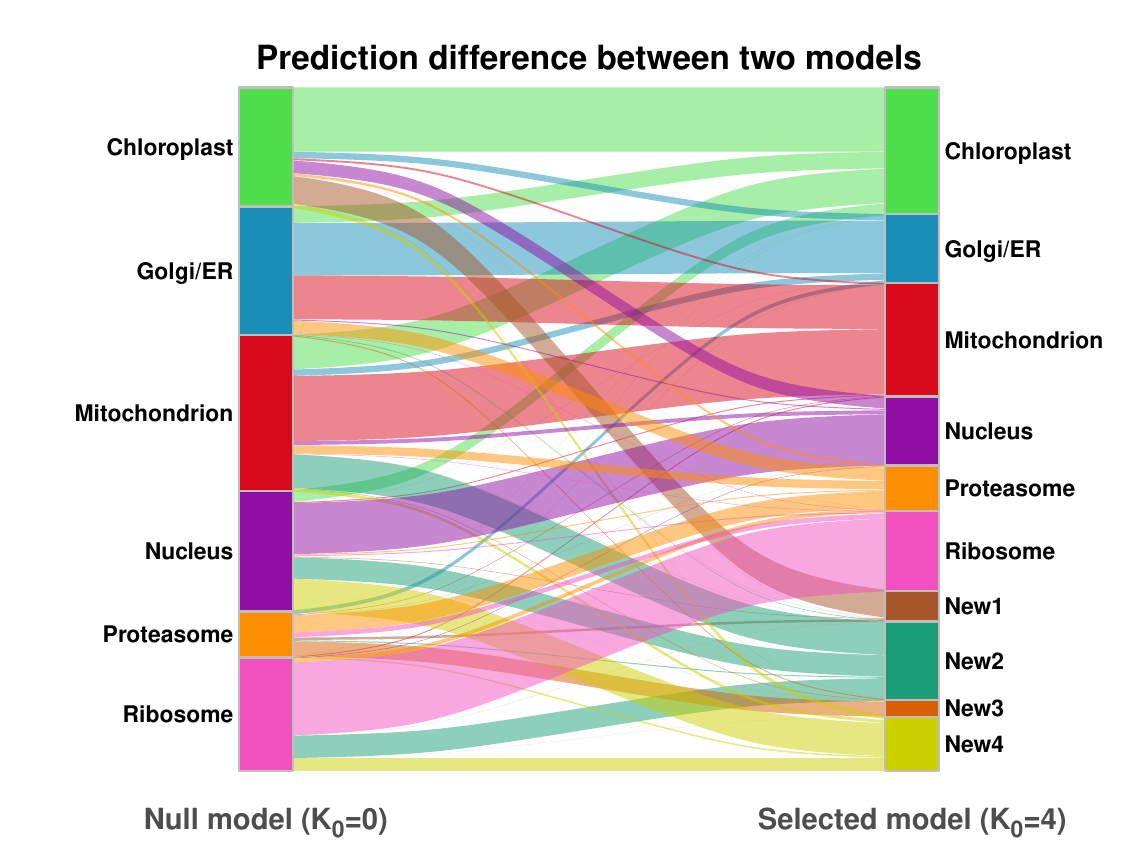}
    \caption{\textbf{Alluvial plot comparing group membership predictions.} 
    This plot visually compares protein membership predictions made by the two FSPmix models (null and selected). The volume of each colored bar flowing from right to left show the relative mix of null model predictions that stem from proteins with a certain membership in the selected model. Most annotated groups with marker proteins have a strong connection between left to right, highlighting the broad-strokes similarity between the two models. }
    \label{fig:alluvial}
\end{figure}

\begin{figure}[H]
    \centering
    \includegraphics[width=\linewidth]{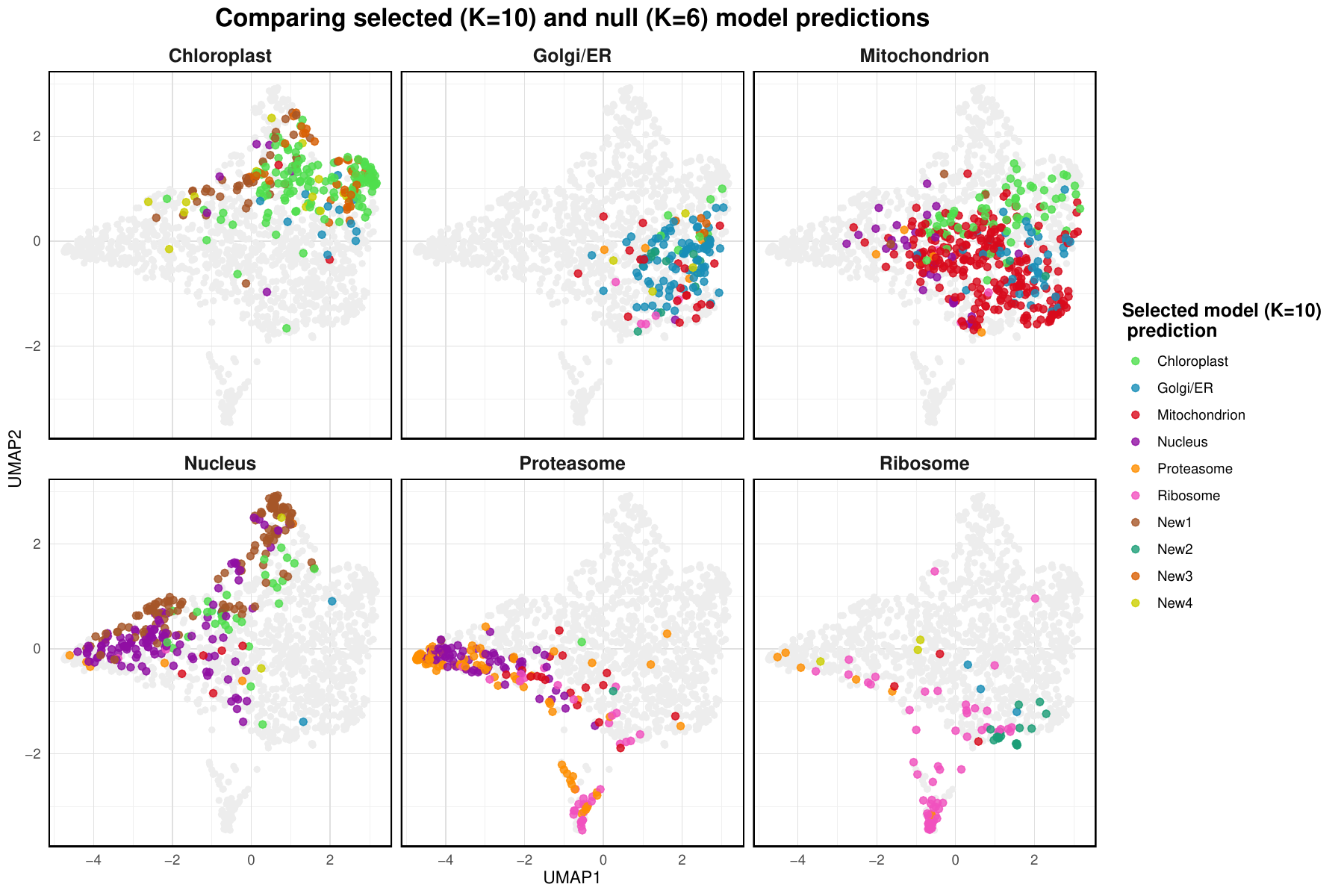}
    \caption{\textbf{Comparing predicted protein groups from two models in UMAP space (continued in Fig~\ref{fig:by-selected}).} 
    This plot shows a detailed view of the predictions made from the two models (null and selected) described in Fig~\ref{fig:umap-summary}. Each panel isolates the proteins predicted to be in one of six group predicted by the null model. Within each panel, the colors of the points in each panel are the ten predicted group memberships from the selected model. Gray points show the full dataset for spatial context. This plot shows that that protein group predictions can differ significantly with and without the FSPmix model's flexibility to discover new groups. }
    \label{fig:by-null}
\end{figure}

\begin{figure}[H]
    \centering
    \includegraphics[width=\linewidth]{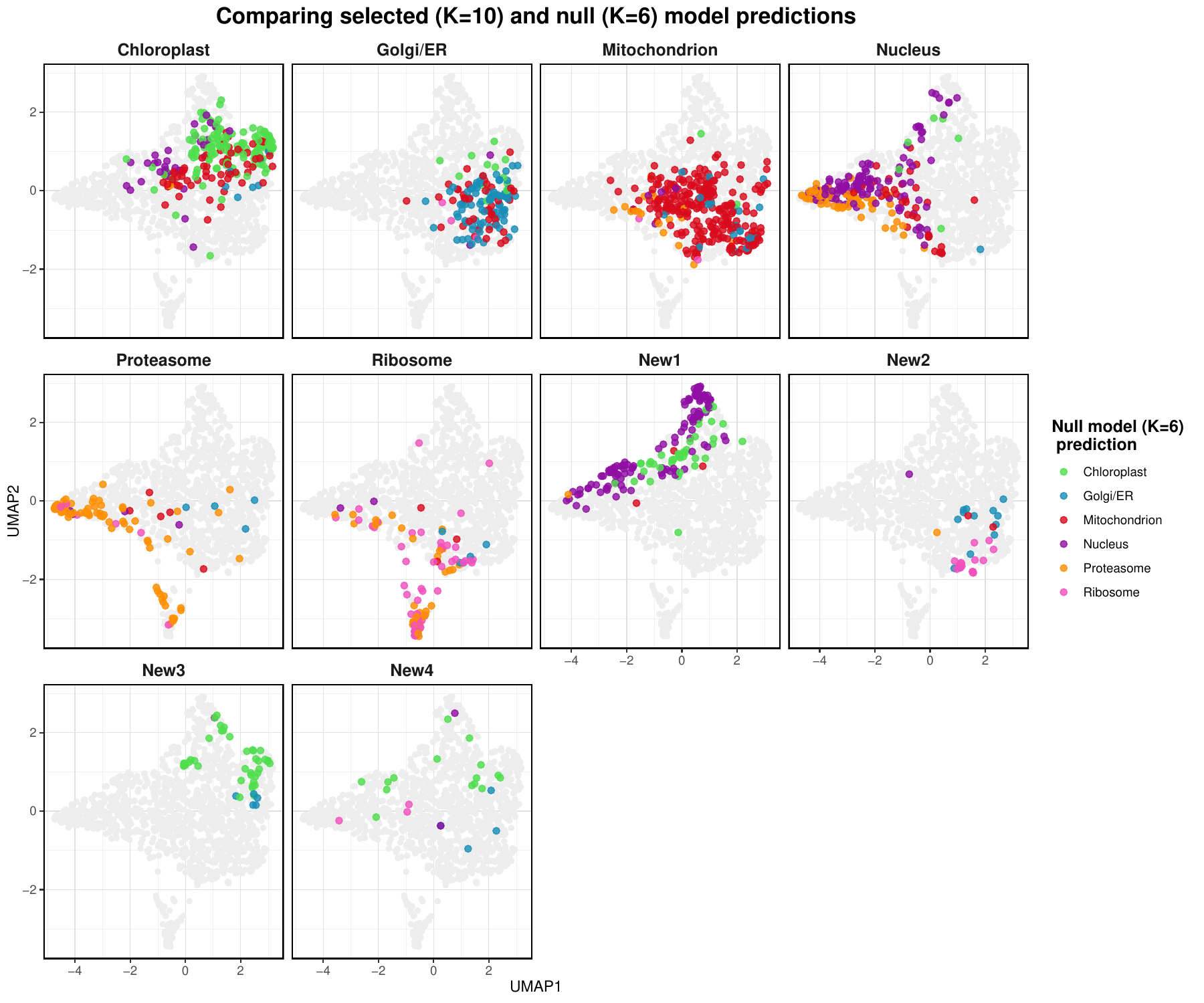}
    \caption{\textbf{Comparing predicted protein groups from two models in UMAP space (continuing from Fig~\ref{fig:by-selected}).} 
    This plot shows a detailed view of the predictions made from the two models (null and selected) described in Fig~\ref{fig:umap-summary}. Each panel isolates the proteins predicted to be in one of ten group predicted by the select model. Within each panel, the colors of the points in each panel are the six predicted group memberships from the null model. Gray points show the full dataset for spatial context. 
    }
    \label{fig:by-selected}
\end{figure}

\begin{figure}[H]
    \centering
    \includegraphics[width=\linewidth]{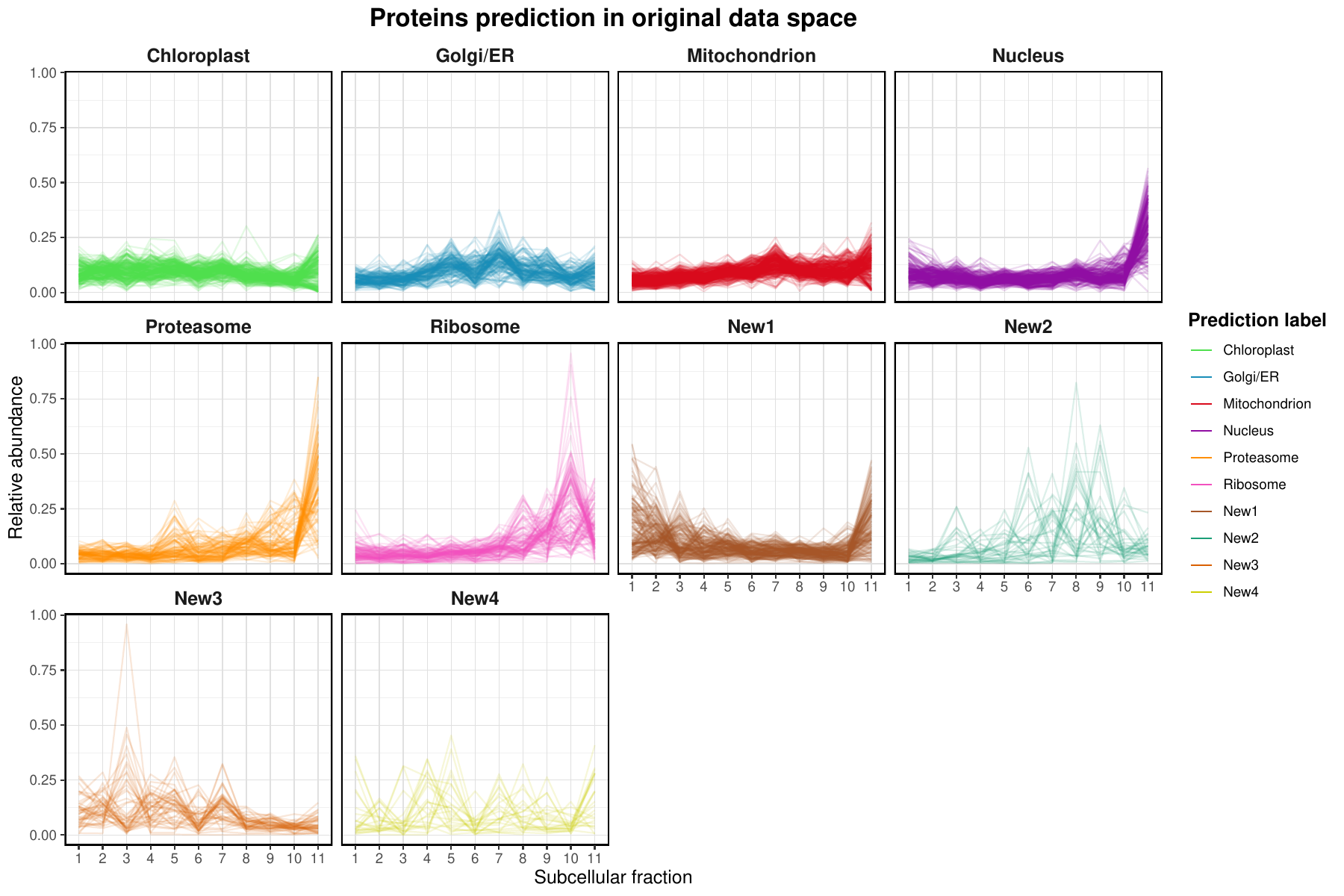}
\caption{Continuing from Fig~\ref{fig:hpb}, {\it all} protein signatures from each of the ten predicted groups (using the ``selected" FSPmix model with $K=10$) are shown.}
\label{fig:original}
\end{figure}

\clearpage

\begin{figure}[H]
    \centering
    \includegraphics[width=\linewidth]{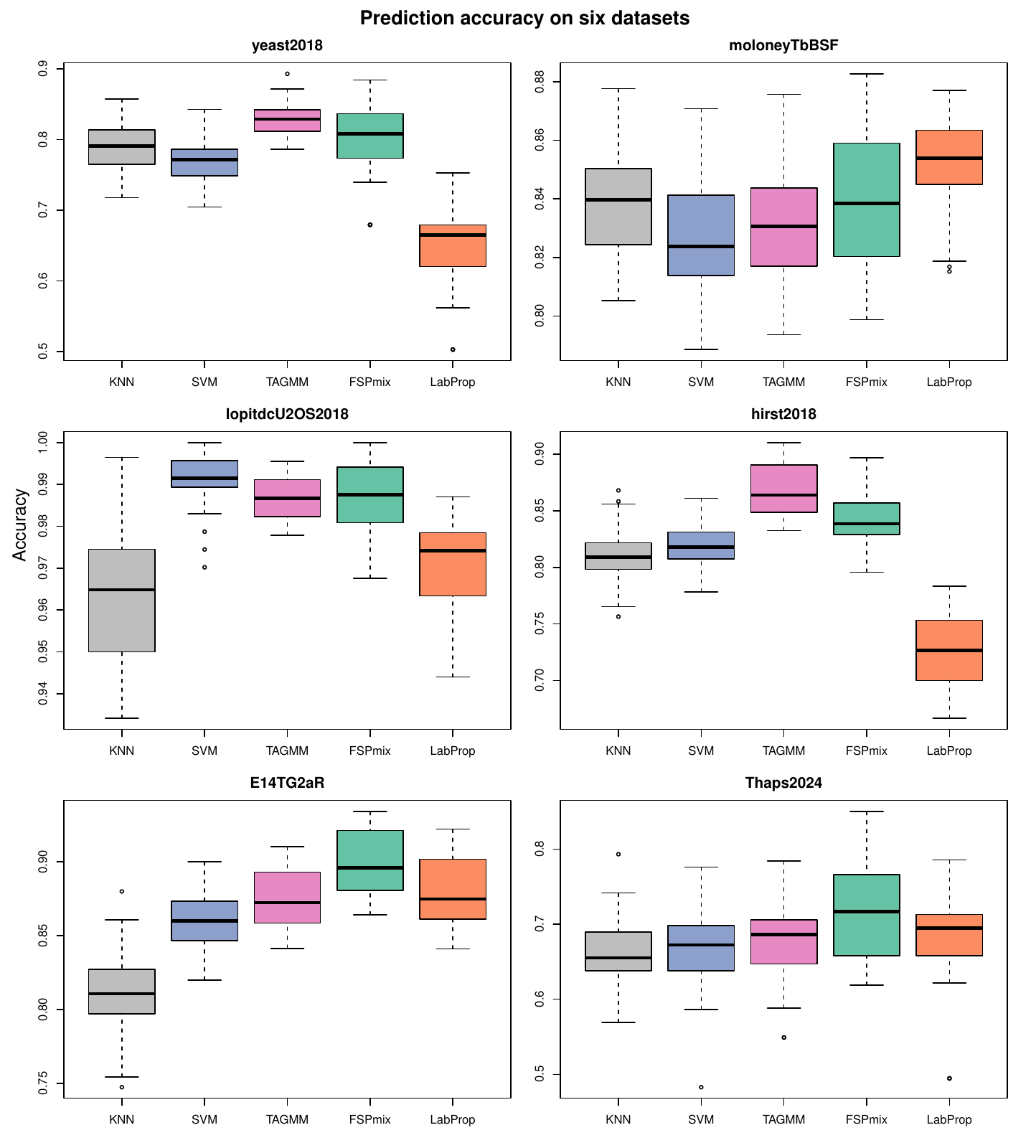}
\caption{
Absolute accuracies from the experiment described in Fig~\ref{fig:acc_comparison} are shown here.
 Each set of boxplots shows the relative group membership prediction accuracy of five semi-supervised models (K-nearest-neighbor (KNN), FSPmix, label propagation (Lab Prop), support vector machines (SVM), and TAGM) from $100$ simulations (described more in detail in Section~\ref{sec:pred_new}. FSPmix outperforms the other four methods especially in the ``low-replicate" datasets containing only a single biological replicate.}
\label{fig:pred_acc}
\end{figure}

\newpage

\bibliographystyle{vancouver}
\bibliography{reference}
\end{document}